\documentclass[fleqn, usenatbib]{mnras}

\usepackage{newtxtext, newtxmath}
\usepackage[T1]{fontenc}

\usepackage{graphicx}
\usepackage{amsmath}
\usepackage{acronym}
\usepackage{dsfont}
\usepackage[separate-uncertainty = true, multi-part-units=single]{siunitx}
\usepackage{amsfonts}
\usepackage{comment}
\usepackage{subcaption}
\usepackage[capitalize]{cleveref}
\usepackage{bm}
\usepackage{listings}
\usepackage{xcolor}
\usepackage{xspace}
\usepackage{siunitx}
\usepackage{multirow}

\DeclareSIUnit{\year}{yr}
\DeclareSIUnit{\day}{d}
\DeclareSIUnit{\hour}{h}
\DeclareSIUnit{\minute}{min}
\usepackage{amstext}

\makeatletter
\@ifpackageloaded{cleveref}{
  \AtBeginDocument{
    \def\ltx@label#1{\cref@label{#1}}
    \def\label@in@display@noarg#1{\cref@old@label@in@display{#1}}
    \def\label@in@mmeasure@noarg#1{%
      \begingroup%
        \measuring@false%
        \cref@old@label@in@display{#1}
      \endgroup}%
  }
}{}
\makeatother

\protected\def\protectedacused{\acused}

\acrodef{LIGO}[LIGO]{Laser Interferometer Gravitational-Wave Observatory}
\acrodef{LHO}[LHO]{\ac{LIGO} Hanford observatory}
\acrodef{LLO}[LLO]{\ac{LIGO} Livingston observatory}
\acrodef{KAGRA}[KAGRA]{KAGRA}\acused{KAGRA}
\acrodef{iKAGRA}[iKAGRA]{initial-phase \ac{KAGRA}}
\acrodef{bKAGRA}[bKAGRA]{baseline-design \ac{KAGRA}}
\acrodef{GEO}[GEO]{GEO\,600 \ac{GW} detector}
\acrodef{aLIGO}{Advanced \ac{LIGO}}
\acrodef{A+}{Advanced+ \ac{LIGO}}
\acrodef{Asharp}[\ensuremath{\text{A}^\sharp}]{\ac{LIGO} \acs{Asharp}}
\acrodef{AdV}{Advanced \acl{Virgo}}
\acrodef{AdV+}{Advanced \acl{Virgo}+}
\acrodef{Virgo}{Virgo}\acused{Virgo}
\acrodef{VirgoNEXT}[Virgo\_nEXT]{Virgo\_nEXT}\acused{VirgoNEXT}

\acrodef{LSC}[LSC]{\acs{LIGO} Scientific Collaboration}
\acrodef{LV}[LV]{\acs{LIGO}--\acs{Virgo} Collaboration\protect\protectedacused{LVC}}
\acrodef{LVC}[LV]{\acs{LIGO}--\acs{Virgo} Collaboration\protect\protectedacused{LV}}
\acrodef{LVK}[LVK]{\acs{LIGO}--Virgo--KAGRA}
\acrodef{IGWN}[IGWN]{International \ac{GWH} Observatory Network}

\acrodef{O1}[O1]{first observing run}
\acrodef{O2}[O2]{second observing run}
\acrodef{O3}[O3]{third observing run}
\acrodef{O3a}[O3a]{first half of the third observing run}
\acrodef{O3b}[O3b]{second half of the third observing run}
\acrodef{O3GK}[O3GK]{observing run}
\acrodef{O4}[O4]{fourth observing run}
\acrodef{O4a}[O4a]{first part of the fourth observing run}
\acrodef{O4b}[O4b]{second part of the fourth observing run}
\acrodef{O4c}[O4c]{third part of the fourth observing run}
\acrodef{O5}[O5]{fifth observing run}

\acrodef{BH}[BH]{black hole}
\acrodef{BBH}[BBH]{binary black hole}
\acrodef{BNS}[BNS]{binary neutron star}
\acrodef{IMBH}[IMBH]{intermediate-mass black hole}
\acrodef{NS}[NS]{neutron star}
\acrodef{BHNS}[BHNS]{black hole--neutron star binaries}
\acrodef{NSBH}[NSBH]{neutron star--black hole binary}
\acrodefplural{NSBH}[NSBH]{neutron star--black hole binaries}
\acrodef{PBH}[PBH]{primordial \ac{BH}}
\acrodef{CBC}[CBC]{compact binary coalescence}

\acrodef{GW}[GW]{gravitational wave\protect\protectedacused{GWH}}
\acrodef{GWH}[GW]{gravitational-wave\protect\protectedacused{GW}}
\acrodef{IFO}[IFO]{interferometer}
\acrodef{SNR}[SNR]{signal-to-noise ratio}
\acrodef{FAR}[FAR]{false alarm rate}
\acrodef{IFAR}[IFAR]{inverse false alarm rate}
\acrodef{FAP}[FAP]{false alarm probability}
\acrodef{PSD}[PSD]{power spectral density}
\acrodef{GR}[GR]{general relativity}
\acrodef{NR}[NR]{numerical relativity}
\acrodef{PN}[PN]{post-Newtonian}
\acrodef{EOB}[EOB]{effective-one-body}
\acrodef{ROM}[ROM]{reduced-order model}
\acrodef{IMR}[IMR]{inspiral--merger--ringdown}
\acrodef{PDF}[PDF]{probability density function}
\acrodef{PE}[PE]{parameter estimation}
\acrodef{CL}[CL]{credible level}
\acrodef{EOS}[EoS]{equation of state}
\acrodef{KLD}[KLD]{Kullback--Leibler divergence}
\acrodef{JSD}[JSD]{Jensen--Shannon divergence}
\acrodef{GCN}[GCN]{general coordinates network}
\acrodef{GWTC}[GWTC]{Gravitational-Wave Transient Catalog}
\acrodef{GWOSC}[GWOSC]{Gravitational Wave Open Science Center}
\acrodef{FFT}[FFT]{fast Fourier transform}

\acrodef{HIQC}[HIQC]{Hierarchical Inference by Quantile Compression}

\acrodef{CWB}[cWB]{coherent WaveBurst}
\acrodef{LAL}[LAL]{\ac{LIGO} algorithm library}

\acrodef{CHRoCC}{central heating radius of curvature correction}
\acrodef{NonSENS}{non-stationary estimation and noise subtraction}

\acrodef{PTA}{Pulsar Timing Array}

\acrodef{MCMC}{Markov chain Monte Carlo}
\acrodef{ESS}{effective sample size}

\acrodef{RS}{rejection sampling}
\acrodef{IS}{importance sampling}
\acrodef{PSIS}{Pareto-smoothed \ac{IS}}
\acrodef{PSRS}{Pareto-smoothed \ac{RS}}

\acrodef{PP}{probability probabilty}
\acrodef{ASD}{amplitude spectral density}
\acrodef{IID}{independent and identically distributed}
\acrodef{KDE}{kernel density estimate}

\newcommand{\soft}[1]{\texttt{#1}}

\newcommand{\GSTLAL}{\soft{GstLAL}\xspace}

\newcommand{\BILBY}{\soft{Bilby}\xspace}

\newcommand{\NUMPY}{\soft{NumPy}\xspace}
\newcommand{\SCIPY}{\soft{SciPy}\xspace}
\newcommand{\MATPLOTLIB}{\soft{Matplotlib}\xspace}

\newcommand{\GWPY}{\soft{GWpy}\xspace}
\newcommand{\DYNESTY}{\soft{Dynesty}\xspace}

\newcommand{\GRAVITYSPY}{\soft{GravitySpy}\xspace}
\newcommand{\OMICRON}{\soft{Omicron}\xspace}
\newcommand{\GLITCHFLOW}{\soft{glitchflow}\xspace}
\newcommand{\ANTIGLITCH}{\soft{antiglitch}\xspace}

\newcommand{\data}{\ensuremath{d}\xspace}
\newcommand{\alldata}{\ensuremath{\vec{d}}\xspace}

\newcommand{\fdata}{\ensuremath{\tilde{d}}\xspace}
\newcommand{\modelG}{\ensuremath{M_{\rm G}}\xspace}
\newcommand{\modelN}{\ensuremath{M_{\rm N}}\xspace}
\newcommand{\parameters}{\ensuremath{\theta}\xspace}
\newcommand{\hyperparameters}{\ensuremath{\vartheta}\xspace}
\newcommand{\hyperparametersfixed}{\ensuremath{\hat{\vartheta}}\xspace}
\newcommand{\basisparameters}{\ensuremath{\zeta}\xspace}
\newcommand{\likelihood}{\ensuremath{\mathcal{L}}\xspace}
\newcommand{\prior}{\ensuremath{\pi}\xspace}
\newcommand{\evidence}{\ensuremath{\mathcal{Z}}\xspace}

\newcommand{\Nbasis}{\ensuremath{N_{\rm basis}}}

\newcommand{\expect}[1]{\left\langle #1 \right\rangle}

\newcommand{\rhomf}{\ensuremath{\rho_{\rm mf}}\xspace}
\newcommand{\lnB}{\ensuremath{\ln B}\xspace}

\newcommand{\pcoinc}{\ensuremath{P_{\rm CG}}\xspace}
\newcommand{\sigmacoinc}{\ensuremath{\sigma_{\rm CG}}\xspace}
\newcommand{\detectorA}{\ensuremath{1}\xspace}
\newcommand{\lambdaA}{\ensuremath{\lambda_{\detectorA}}\xspace}
\newcommand{\detectorB}{\ensuremath{2}\xspace}
\newcommand{\lambdaB}{\ensuremath{\lambda_{\detectorB}}\xspace}
\newcommand{\Tsep}{\ensuremath{\delta t}\xspace}
\newcommand{\Tspan}{\ensuremath{T}\xspace}

\newcommand{\Nseg}{\ensuremath{N_s}\xspace}
\newcommand{\Msamples}{\ensuremath{M}\xspace}
\newcommand{\Quantiles}{\ensuremath{Q}\xspace}

\newcommand{\changed}[1]{\textcolor{black}{#1}}
\newcommand{\minorchanged}[1]{\textcolor{black}{#1}}

\title[Measuring the glitch rate]{Measuring the rate of glitches in interferometric gravitational wave detectors with a hierarchical Bayesian model}

\author[G.~Ashton et al.]{
Gregory Ashton$^{1,2}$\thanks{E-mail: gregory.ashton@ligo.org},
Colm Talbot$^{3}$,
Andrew Lundgren$^{4,5}$,
Ann-Kristin Malz$^{2}$,
Joseph Areeda$^{6}$
\\
$^{1}$Mathematical Sciences, University of Southampton, Southampton SO17 1BJ, United Kingdom\\
$^{2}$Department of Physics, Royal Holloway, University of London, Egham Hill, Egham, TW20 0EX, United Kingdom\\
$^{3}$Department of Physics, Princeton University, Princeton, NJ 08544, USA\\
\minorchanged{$^{4}$Institut de F\'{\i}sica d'Altes Energies, E-08193 Barcelona, Spain}\\
\minorchanged{$^{5}$Instituci\'{o} Catalana de Recerca i Estudis Avan\c{c}ats, E-08010 Barcelona, Spain}\\
$^{6}$California State University Fullerton, Fullerton, CA 92831, USA\\
}

\pubyear{2026}

\begin{document}
\label{firstpage}
\pagerange{\pageref{firstpage}--\pageref{lastpage}}
\maketitle

\begin{abstract}
Ground-based gravitational wave detectors are now routinely surveying the dark Universe, finding hundreds of \minorchanged{collisions between compact objects}.
However, terrestrial non-Gaussian noise artefacts, commonly known as glitches, reduce the sensitivity to signals and can overlap signals, producing biased astrophysical inferences.
We introduce a hierarchical Bayesian model to measure the glitch rate, which improves upon existing trigger-counting methods in its capacity to measure the rate down into the low signal-to-noise regime without contamination from the Gaussian noise background, provided the population is accurately modelled.
\changed{The framework accommodates any glitch model, and measures the rate with respect to the model chosen: here we use the \ANTIGLITCH model, so the rate inferred is that of short-duration glitches rather than of all glitches.}
The methodology builds on standard hierarchical inference, but includes several novel features: \minorchanged{hierarchical inference with quantile compression (HIQC), a generic approximation for the recycled hyperlikelihood,} and a time-domain rate estimated by fitting basis functions.
We validate the methodology using simulated data with injected glitches and then
apply it to data from the fourth LIGO-Virgo-KAGRA observing run, demonstrating time-resolved inferences of the glitch rate over a \qty{24}{\hour} period.
The inferred glitch rate is consistent with estimates from trigger counts, but \minorchanged{requires no} arbitrary threshold and provides a more fine-grained view of the temporal behaviour.
Finally, we demonstrate how our individual-detector rate estimates can be transformed into a coincident glitch probability and utilise this to \changed{provide evidence} that the retracted gravitational-wave candidate GW230630\_070659 is likely a pair of coincident glitches.

\end{abstract}


\section{Introduction}
\label{sec:introduction}

Since the inception of the \aclu{LIGO} \citep[LIGO:][]{LIGOScientific:2014pky}, it has been known that non-Gaussian transient noise triggers are an unwanted feature of the interferometric strain and occur at a rate of a few times per hour \citep{Abramovici:1992ah}.
These noise sources have now come to be known as \emph{glitches}
\citep[see][for a recent review]{Davis:2022dnd} and they are also known
\citep{Virgo:2022ysc, Akutsu:2025ubp} to affect the
Virgo \citep{VIRGO:2014yos} and KAGRA \citep{KAGRA:2020tym} detectors.
In some cases, noise investigations reveal couplings between the glitches observed in the strain and environmental channels \citep{Nuttall:2015dqa,Berger:2018ckp,McIver:2019hqm,LIGO:2021ppb}, enabling either mitigation of the source or vetoes to reduce the rate of false positives \citep[see, e.g.][]{Kotter:2003nq,Hanna:2006ub}.
While this has reduced the prevalence of some glitch classes, many have no environmental coupling, and as the Gaussian noise background is reduced, a constant background of glitches remains in the data \minorchanged{\citep[see, e.g.][]{LIGO:2024kkz}}.

Glitches have two significant impacts. First, they reduce the sensitivity to astrophysical sources by increasing the rate of false positives and hence requiring a stricter threshold for detection.
Second, they can overlap astrophysical sources, biasing the inferred properties
\citep{Udall:2025bts, Hourihane:2025vxc}.
Therefore, significant effort is undertaken to investigate sources and understand the statistical properties of glitches \citep{Blackburn:2008ah}.
In the most recent analysis of data from the first part of the fourth observing run (O4a), it was found that the rate of glitches was in the region of tens per hour \citep{LIGO:2024kkz} and similar rates were also reported during the third observing run \citep{KAGRA:2021vkt}.

Measuring the glitch rate is non-trivial.
The standard approach is to analyse the strain data and auxiliary channels to find \emph{triggers} using the \OMICRON software \citep{Robinet:2020lbf}.
Applying a time-frequency decomposition, \OMICRON is applied in batches to all observed data, producing a list of triggers with corresponding measures of the \ac{SNR} and other properties.
The rate can then be estimated as the number of events within the analysis duration, assuming Poissonian statistics.
However, as with searching for any transient signal, in the absence of any glitches, Gaussian noise fluctuations will produce a distribution of triggers in \ac{SNR} space; if the right-hand tail of this distribution overlaps the distribution of \acp{SNR} produced by glitches, then the two cannot be unambiguously separated.

\begin{figure}
    \centering
    \includegraphics[width=\linewidth]{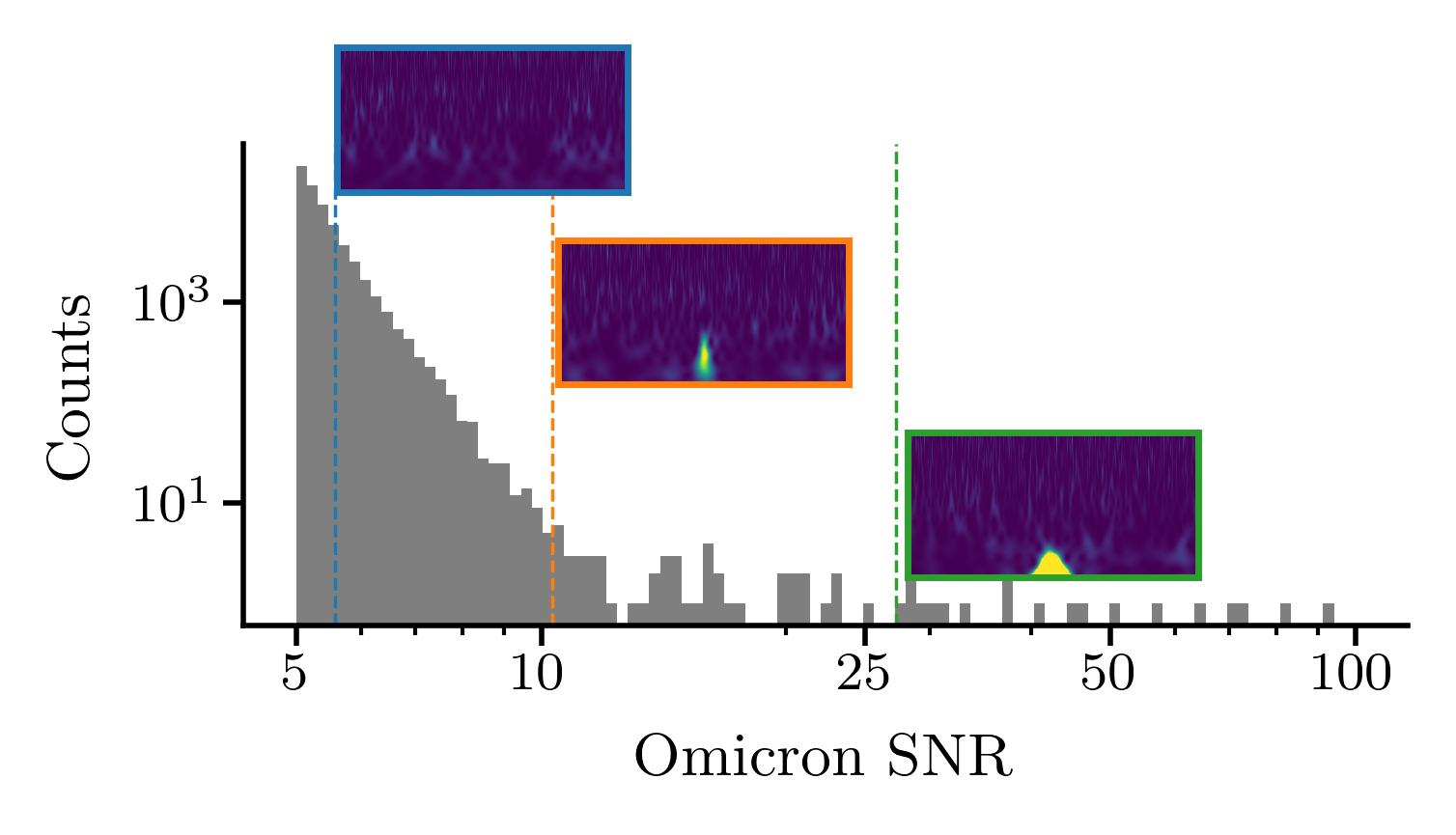}
    \caption{A histogram of the counts of \OMICRON trigger \acp{SNR} over one day of data (2023-08-18) in the LIGO Livingston detector during O4a, along with time-frequency spectrograms \citep[created using \GWPY:][]{Macleod:2021goi} of triggers selected from representative regions of the distribution: vertical dashed lines denote their \ac{SNR}.
    The spectrograms cover a time span of \qty{1}{\second} and a frequency range from \qty{20}{\Hz} to \qty{1}{\kilo\Hz} \changed{and share the same color scale for comparison}.}
    \label{fig:glitches}
\end{figure}

For the \ac{LVK} detectors, the distribution of \OMICRON \acp{SNR} for glitches does overlap the noise distribution.
As an example, in \cref{fig:glitches} we show the distribution of \OMICRON trigger \acp{SNR} from \ac{LLO} over a day in O4a down to a minimum \ac{SNR} of \num{5}.
From the time-frequency spectrograms of representative points in the distribution, we can observe that at the minimum \ac{SNR}, the data is consistent with Gaussian noise.
Meanwhile, triggers with larger \acp{SNR} show distinct glitch behaviour.
From the \ac{SNR} distribution itself, there is no obvious threshold to disentangle the glitches from the Gaussian noise fluctuations.

\begin{figure}
    \centering
    \includegraphics[width=\linewidth]{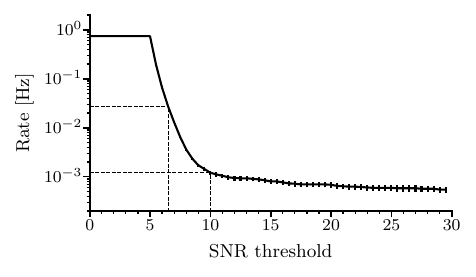}
    \caption{The estimated Poissonian glitch rate (and error) from \minorchanged{\OMICRON} triggers for one day of data during O4a as a function of the \minorchanged{\OMICRON} SNR threshold. Dashed lines mark the thresholds of \num{6.5} and \num{10} that are often used to determine the glitch rate.}
    \label{fig:omicron_rate}
\end{figure}

That the distributions overlap leads to a difficulty in measuring the rate of glitches: namely, a threshold must be selected to count glitches in a given time period.
In \citet{LIGO:2024kkz}, \ac{SNR} thresholds of \num{6.5} and \num{10} were used and these are common choices \citep[see, e.g. ][]{KAGRA:2021vkt}.
Taking the day of O4a triggers from \cref{fig:glitches}, we extend the notion of measuring the rate at a fixed threshold, and in \cref{fig:omicron_rate}, we plot the rate against the choice of \ac{SNR} threshold.
We see that above a threshold of \num{10}, the rate is reasonably stable, below $10^{-3}$~Hz.
However, below an SNR of \num{10}, the rate markedly increases by orders of magnitude.
This highlights the difficulty: choosing a conservative threshold will necessarily lead to an underestimate of the glitch rate, while choosing a more liberal threshold (e.g., an SNR of \num{6.5}), the glitch rate may be overestimated due to false positives arising from Gaussian noise.
Nevertheless, glitch rates based on an \ac{SNR} threshold are the standard method used by the \ac{LVK} to report glitch activity \citep{LIGO:2024kkz}.
Moreover, the set of glitches obtained also forms the basis of further investigation.
For example, \citet{Costa:2025pmd} investigates the consistency of the measured glitch time with a Poisson distribution while \citet{Ferreira:2024gzh} uses a clustering algorithm to understand the temporal behavior.

This work aims to introduce an alternative method for measuring the glitch rate using a hierarchical Bayesian framework.
In part, this was motivated by work searching for the astrophysical background of merging compact binary coalescence \citep{Smith:2017vfk}.
In short, the method centres on the idea of applying full Bayesian parameter estimation to the full stream of data.
Compared to the background search of \citet{Smith:2017vfk}, our method differs in that: (i) we do not need to use coherence to filter out glitches (since we aim to measure the rate of glitches directly); (ii) we consider data from a single detector at a time; and (iii) we do not use an astrophysical model of \acp{CBC} signals, but a generative model for glitches.
We will use the \ANTIGLITCH model developed by \citet{Bondarescu:2023jcx}.
As such, our analysis will be sensitive only to glitches for which the \ANTIGLITCH model is effective.
For glitches where it is not a good fit (in the sense that the measured \ac{SNR} is low), we will be insensitive, and hence these will implicitly not be counted in our rate estimate. 
In \citet{Bondarescu:2023jcx}, known glitches labelled by \GRAVITYSPY \citep{Zevin:2016qwy} (which internally selects from the \OMICRON triggers) were analysed with the \ANTIGLITCH model.
They found that the \ANTIGLITCH model was a good general-purpose model for a wide variety of short-duration glitches.
Therefore, we expect our method to provide a good estimate for short-duration glitches, but for long-duration glitches, a separate model may be needed.
We will leave the development of this to future work, but the algorithm can naturally generalise by using a mixture-model likelihood with multiple (possibly overlapping) glitch models.

This paper is organised as follows.
In \cref{sec:methodology}, we introduce the methodology, starting with the standard counting-based approach, then developing the levels of the hierarchical model and demonstrating how it can be extended to time-dependent and population-dependent analysis.
We then perform a set of validation studies in \cref{sec:validation}, simulating glitches in Gaussian noise and testing whether our method can recover their statistical properties.
In \cref{sec:results}, we apply the methodology to open data from LIGO during O4a, studying a day of data
before going on to look at a specific astrophysical event in \minorchanged{\cref{sec:app}}.
Finally, in \cref{sec:discussion} we provide a discussion about the future applicability of the method.

\section{Methodology}
\label{sec:methodology}

\subsection{Counting-based rate estimation}
\label{sec:counting}
To begin our discussion of the methodology, we first review the standard approach to rate estimation by counting the number of \OMICRON triggers above a given \ac{SNR} threshold.
Underlying this approach is the assumption that glitches are produced by some Poisson process with a rate per unit time, $\lambda$, such that the probability mass function for $k$ events in an interval $\Delta T$ is
\begin{align}
    P(k)=\frac{(\lambda \Delta T)^k e^{-\lambda \Delta T}}{k!}\,.
\end{align}
Given $n$ observed events, estimators of the rate and its standard error can then be calculated:
\begin{align}
    \hat{\lambda} = \frac{n}{\Delta T}\,.
    \label{eqn:lambda_hat}
\end{align}
and
\begin{align}
    \sigma_{\hat{\lambda}} = \frac{\hat{\lambda}}{\sqrt{n}}=\frac{\sqrt{n}}{\Delta T}\,.
    \label{eqn:lambda_hat_err}
\end{align}
(We note that \OMICRON triggers can be \minorchanged{clustered, with} multiple triggers corresponding to the same glitch.
Therefore, when calculating the rate in \minorchanged{practice}, we apply a simple clustering approach, keeping only the highest-\ac{SNR} trigger within a \minorchanged{given window.)}

This straightforward approach is simple and intuitive, but it assumes that the rate itself is constant over the duration $\Delta T$; for gravitational-wave detectors, this is unlikely to be true if $\Delta T$ is longer than a few tens of minutes.
\changed{We emphasise that this is a statement about the glitch-generating process, not about our ability to measure it: the state of the instrument, including its alignment, thermal condition, and the anthropogenic and environmental activity around it, evolves on timescales of tens of minutes, and it is therefore likely to expect the process generating the glitches to evolve with it.
}
When the rate is not constant, the estimator will produce a time-averaged rate, which can be biased, the uncertainty will typically be underestimated, and the approach ignores trends.
Therefore, it is often preferable to break a long data set into short, possibly overlapping, windows and estimate the rate within each window.
For example, in \citep{LIGO:2024kkz}, windows of \qty{1}{day} and \qty{1}{hour} were used, while \citet{KAGRA:2021vkt} used bins of \qty{2048}{\second}. 
This windowing approach has the advantage of enabling the study of trends.
However, reducing $\Delta T$ increases the uncertainty on the estimate: from \cref{eqn:lambda_hat_err} we can invert to get an approximate rule of thumb for the duration required to measure at a relative accuracy of $R$ for a given rate:
\begin{equation}
    \Delta T \approx \qty{100}{minutes}\,
    \changed{\left(\frac{10\%}{R}\right)^2}
    \left(\frac{\qty{1}{\per \minute}}{\lambda}\right)\,.
    \label{eqn:rule_of_thumb}
\end{equation}
Taking $\Delta T$ to be shorter than this, the relative error increases.
In the extreme cases where $\Delta T$ is so small that no events occur, the method breaks down, providing an estimated rate and uncertainty of zero.

\changed{The two considerations that set the window are therefore in tension.
The timescale on which $\lambda$ genuinely varies is a property of the instrument and is not ours to choose, while the precision available in any window is set by the number of glitches that fall within it.
\Cref{eqn:rule_of_thumb} quantifies the conflict: at $\lambda\sim\qty{1}{\per\minute}$, measuring the rate to \qty{10}{\percent} requires a window of \qty{100}{\minute}, which is already longer than the timescale on which we expect the rate to change.
Choosing a window short enough to resolve the variation thus costs precision, and choosing one long enough for precision may average the variation away.
}

A standard Bayesian approach to maintain consistent uncertainty estimates when $n$ is small is to calculate the posterior distribution on $\lambda$ using the conjugate prior to the Poisson distribution, which is the Gamma distribution.
\changed{We adopt the shape--rate parameterisation throughout.
Choosing a Gamma prior with shape parameter \num{1} and rate parameter \num{0}, i.e. $\pi(\lambda) = \mathrm{Gamma}(1, 0) \propto \lambda^{1-1}e^{-0\cdot\lambda} = 1$, recovers the (improper) uniform prior on $\lambda$; this is the degenerate limit of the Gamma family as the rate tends to zero, but the resulting posterior is proper for any $\Delta T>0$.}
Under this assumption, the posterior distribution is another Gamma distribution:
\begin{align}
   p(\lambda | n) = \minorchanged{\mathrm{Gamma}}(n+1, \Delta T)\,, 
\end{align}
which has a mean $(n+1)/\Delta T$ that recovers the standard estimator, \cref{eqn:lambda_hat}, when $n\gg 1$.
However, when $n$ is zero, the uncertainty is finite (the inverse of $\Delta T$), and an upper limit can be derived from the Gamma distribution.
This Gamma-posterior approach is just as easy to compute as the simple Poisson rate estimate, but naturally models the uncertainty as the number of observed events tends to zero.
From hereon, we will refer to these two approaches as the Poisson and Gamma counting-based approaches, respectively.

\subsection{Hierarchical Bayesian Model}
The goal of our hierarchical framework is to infer the statistical properties of the glitch-generating process from the strain data.
We separate this inference into two levels and describe them in general terms in this section.

The Level-II inference infers the posterior distribution of some population-level hyperparameters \hyperparameters: 
\begin{align}
    p(\hyperparameters | \alldata) \propto \likelihood(\alldata | \hyperparameters)\pi(\hyperparameters)\,,
    \label{eqn:hyperposterior}
\end{align}
where $\alldata$ is a timeseries of data from a single interferometer which is sufficiently long as to include multiple glitches, $\likelihood(\alldata | \hyperparameters)$ is the hyperlikelihood of the data given the population hyperparameters, and $\pi(\hyperparameters)$ is the prior distribution on the hyperparameters.

Meanwhile, Level-I inference operates on short segments of data.
Specifically, we segment $\alldata$ into a set of \Nseg (possibly overlapping) strain segments: $\vec{\data} = \left\{ \data_i\,\forall\,i\,\in [0, N_s-1] \right\}$.
The duration of $\data_i$ should be sufficiently short such that at most they contain a single glitch.
(We recognise that in fact glitches can be highly non-Poissonian and clustered: we will return to this later in \minorchanged{\cref{sec:td}.)}
For each \minorchanged{Level-I} analysis, we analyse the data under a predictive glitch model \modelG with an associated set of parameters \parameters and a fixed choice of the population hyperparameters, which we denote as \hyperparametersfixed.
Applying computational Bayesian inference, this analysis produces an estimate of the evidence
\begin{align}
    \evidence(\data_i | \hyperparametersfixed, \modelG) = \int_\parameters
    \likelihood(\data_i | \parameters, \modelG) \prior(\parameters | \hyperparametersfixed, \modelG) \, d\parameters\,,
    \label{eqn:evidence}
\end{align}
with an associated uncertainty and an estimate of the posterior
\begin{align}
    p(\parameters | \data_i, \hyperparametersfixed, \modelG) \propto \likelihood(\data_i | \parameters, \modelG) \prior(\parameters | \hyperparametersfixed, \modelG)\,,
    \label{eqn:posterior}
\end{align}
in the form of a set of $\Msamples_i$ posterior samples $\parameters_i^\ell \sim p(\parameters | \data_i, \hyperparametersfixed, \modelG)$ where the subscript indexes the data set and the superscript indexes the sample number $\ell \in [1, \Msamples_i]$.

\changed{To be explicit about which quantities vary at which level: the glitch parameters \parameters are sampled over in the Level-I inference and are marginalised out in \cref{eqn:evidence}, so they do not appear at Level-II.
The hyperparameters \hyperparameters are the only quantities sampled over at Level-II; they are held fixed at the reference values \hyperparametersfixed throughout the Level-I inference, which is what makes the Level-I analyses reusable.
The hat therefore denotes ``fixed at the Level-I reference value'', not an estimator: \hyperparametersfixed is a choice we make before running any Level-I analysis, and its role is to define the prior $\prior(\parameters | \hyperparametersfixed)$ under which the Level-I evidences and posteriors are computed.
The purpose of the recycling step in \cref{sec:pm} is precisely to convert results obtained at \hyperparametersfixed into the quantities required at an arbitrary \hyperparameters, without re-running the Level-I inference.}

After analysing all \Nseg data segments, we construct the hyper-likelihood of \cref{eqn:hyperposterior} using a two-component mixture model \changed{applied \emph{to each segment}}: every $\data_i$ either includes a glitch (\modelG) or contains only coloured Gaussian noise (\modelN).
\changed{The mixture is therefore taken inside the product over segments:}
\begin{align}
    \changed{\likelihood(\alldata | \hyperparameters) =
    \prod_{i=1}^{\Nseg}
    \Big[
    P(\modelG| \hyperparameters)\,\evidence(\data_i | \hyperparameters, \modelG)}
    \changed{+ \big(1 - P(\modelG| \hyperparameters)\big)\,\evidence(\data_i | \modelN)
    \Big]\,,}
    \label{eqn:hyperlikelihood}
\end{align}
where $P(\modelG | \hyperparameters)$ is the probability that \changed{a segment} contains a glitch, $P(\modelN)=1-P(\modelG | \hyperparameters)$ is the probability it contains only Gaussian noise, and $\evidence(\data_i | \modelN)$ is the evidence for the \modelN model.
\changed{The product over segments in \cref{eqn:hyperlikelihood} assumes the segments provide statistically independent measurements.
For the overlapping segments we use in practice this is not strictly true, and we return to the consequences in \cref{sec:levelI}.}

Finally, we assume the glitch-generating mechanism is Poissonian and define $\lambda \in \hyperparameters$ as the per-unit-time rate parameter such that the probability of a glitch in a segment of duration $\delta t$ can be calculated from
\begin{align}
    P(\modelG | \hyperparameters) = \lambda \delta t \exp\left(-\lambda \delta t\right)\,
\end{align}
and therefore $P(\modelN) = 1 - P(\modelG | \hyperparameters)$.

Now that we have outlined the general structure of the hierarchical model, we discuss the details of the Level-I and Level-II inference in the following sections.


\subsection{Level-I inference}
\label{sec:levelI}
The general principle of the Level-I inference follows standard parameter estimation methodology for transient astrophysical signals \citep[see, e.g.][]{2015PhRvD..91d2003V, Thrane:2018qnx},
except that, unlike an astrophysical source, the signal is modelled directly as the single-detector strain response. 
That is, we model the observed time-series strain data \data as the additive sum of a coloured Gaussian noise process and a deterministic glitch model $\mu(\parameters)$, where \parameters denote the glitch parameters.

\emph{Likelihood} ---
Under these assumptions, we will use the Whittle likelihood such that under model \modelG, the log-likelihood is given by
\begin{align}
    \log \likelihood(\fdata_i | \parameters, \modelG)
    \propto -\frac{1}{2}\expect{\fdata-\tilde{\mu}(\parameters), \fdata - \tilde{\mu}(\parameters)}\,,
\end{align}
where $\fdata_i$ is the \ac{FFT} of the $i$th data segment, $\tilde{\mu}(\parameters)$ is the frequency-domain prediction of the glitch model,
\begin{align}
    \expect{a,b} \equiv \frac{4}{T} \sum_{j} \mathbb{R} \left(\frac{a_j^* b_j}{P_j}\right)\,
    \label{eqn:inner_product}
\end{align}
is the noise-weighted inner product, and $P_j$ is the estimated \ac{PSD}.
To estimate the \ac{PSD}, we will utilise Welch's method \citep{welch1967use} with a 50\% overlap, median averaging, and using \qty{64}{s} data preceding $\data_i$.

\emph{Parameterised glitch model} ---
For this work, we use the \ANTIGLITCH model \citep{Bondarescu:2023jcx} which is defined in the frequency domain and built around a Gaussian function of the log-frequencies.
We define it here with a modified notation.
First, we define the base Gaussian for the $j$th frequency bin as
\begin{align}
    \tilde{\mu}_j^{(0)} = \exp\left(-\frac{\gamma}{2}\left(\log{f_j} - \log{f}\right)^2\right)\,,
\end{align}
where $\log{f}$ and $\gamma$ are the mean and precision of the Gaussian (in \citet{Bondarescu:2023jcx}, $\gamma$ is called the inverse bandwidth squared, but here we \changed{adopt the statistical convention in which the precision is the reciprocal of the variance}).  \changed{Since $\log f_j - \log f = \log(f_j / f)$ is dimensionless, $\gamma$ is likewise dimensionless.}
Then the full model adds an amplitude $A$, central time $t_c$, and phase offset $\varphi$:
\begin{align}
    \tilde{\mu}_j = \mathcal{N}^{-1} A \exp\left(2i \varphi - 2\pi i f_j t_c\right) \tilde{\mu}_j^{(0)} \,,
\end{align}
where
\begin{align}
    \mathcal{N} = \sqrt{\sum_j \frac{|\tilde{\mu}_j^{(0)}|^2}{P_j}}\,,
\end{align}
is a normalization factor.
In total, the \ANTIGLITCH model has five parameters $\theta=\{f, A, \gamma, \varphi, t_c\}$ and we note that, unlike astrophysical signals, where the two polarisations must be projected onto the antennae pattern functions, here $\tilde{\mu}(\parameters)$ directly models the detector response.

 The \ANTIGLITCH model is one of many glitch models that could be applied \citep[see, e.g.][]{2023ApPhL.122i4103U,2024CQGra..41m5015S}.
Moreover, \ANTIGLITCH can be extended by modelling glitches as a sum of \ANTIGLITCH terms, capturing multiple components.
In future work, it would be interesting to extend the hierarchical analysis to include these alternative glitch models by including mixtures of glitch types.
However, in this work, we restrict ourselves to searching for glitches that can be modelled by the \ANTIGLITCH model.
This will limit our sensitivity, and any population statements will apply only to the population of glitches for which the \ANTIGLITCH model performs reasonably well.
However, it will enable us to set out the basic structure without overcomplication.

\emph{Prior} --- We use uniform priors on the five glitch parameters, as given in \cref{tab:priors}.
We note that these differ from those proposed in \citet{Bondarescu:2023jcx}, but cover a similar range.

\begin{table}
    \centering
    \begin{tabular}{c|l|r}
         Parameters & Prior & Units \\\hline
         $f$ & $\textrm{Unif}(15, 256)$ & \unit{\Hz} \\
         $A$ & $\textrm{Unif}(0, 1000)$ & --- \\
         $\gamma$ & $\textrm{Unif}(0.01, 20)$ & \changed{---} \\
         $t_c$ & $\textrm{Unif}(t_0 - \Delta t/2, t_0 + \Delta t / 2)$ & \unit{\second} \\
         $\varphi$ & $\textrm{Unif}(-\pi, \pi)$ & \unit{\radian} \\
    \end{tabular}
    \caption{Prior distributions used in the Level-I analysis. Here $t_0$ is the middle of the segment $d_i$ and $\Delta t=\qty{1.1}{\second}$.}
    \label{tab:priors}
\end{table}

\emph{Explicit marginalization} ---
With the choice of the \ANTIGLITCH model, the Whittle likelihood is amenable to explicit marginalisation of the phase and central time \citep{Thrane:2018qnx}.
We implement this with a modified version of the Whittle likelihood implemented in \BILBY, including reconstruction of the phase and time parameters after sampling.
We find that this significantly reduces the computational overhead of the analysis.

\emph{Computational inference} ---
We use \BILBY \citep{Ashton:2018jfp,Romero-Shaw:2020owr} to perform the Level-I inference, using the nested sampling algorithm implemented in \DYNESTY \citep{Speagle:2019ivv}.

\emph{Segment setup and timing model}---
For experiments presented in this work, we will use streams of uninterrupted data \alldata with a duration $T$ that is hours long or more.
We intend for our Level-I analyses to use \qty{4}{\second} segments of data with a central-time prior window of \qty{1}{\second} centred in the middle of the segment.
To ensure we identify all glitches in the full data set \alldata, we define the start time of data segment $\data_i$ to be
\begin{align}
    t_i = t_0 + i \times \qty{1}{\second}
\end{align}
where $t_0$ is the start time of the full data set.
In this way, to cover a duration of $T$ \unit{\second} of data, we need to analyse $T$ segments.
This setup over-covers in two ways.
First, the analysis of the $i$th segment includes \qty{3}{\second} of data from the adjacent segments.
Second, we choose a time-prior span of $\Delta t=\qty{1.1}{\second}$ (cf. \cref{tab:priors}), which results in an overlap at the edges.
This overcovering is done intentionally, as it was found that a few glitches occur at the boundary of the pre-defined segments and can be missed if we don't overcover.
With over covering, we find that adjacent segments can both find the same glitch.

If left unaddressed, the overcovering will result in an overestimate of the glitch rate, since both segments are treated as independent observations of a glitch.
Therefore, to mitigate this, before using the Level-I inference results, we cluster adjacent triggers with positive Bayes factors and discard all but the maximum.
\changed{
This will result in an underestimate, as the number of noise segments is artificially reduced, but we expect the effect to be small except when the glitch rate is unusually large.
We caution, however, that this is a systematic we have not quantified: the clustering step, and more generally the correlations introduced by the \qty{1}{\second} stepping of \qty{4}{\second} segments, are not captured by the independent-segment product of \cref{eqn:hyperlikelihood}.
A dedicated study, in which the recovered rate is compared against the known injected rate as a function of the segment overlap, would quantify the resulting bias, and improved methods will be required to handle these edge cases and produce estimates that are unbiased by construction.
We leave both to future work.}

The two primary computational costs associated with the \minorchanged{Level-I} analyses are the read-in of the data and the stochastic sampling.
Working on the LIGO data grid, where data is stored locally, the read-in time of the data is fast, just a few seconds, but this can be slower if the analysis first needs to download the data.
Meanwhile, for both the simulated validation data and the observed data from the LIGO detectors (see \cref{sec:timing}), we find that the sampling time of segments consistent with Gaussian noise (as inferred from the Bayes factor) to have 90\%  upper interval of $\approx$~\qty{40}{\second} per \qty{1}{\second} of data.
(These timings were computed on the available mixture of nodes of the Caltech computing cluster, part of the LIGO Data Grid; while faster nodes are available, we present this as a realistic timing model estimate based on available resources.)
Segments that contain glitches can take up to ten times longer.
However, for typical rates, where glitches are rare, this means that the overall computation time associated with the \minorchanged{Level-I} analysis is dominated by the Gaussian noise segments.
Neglecting the data read-in time, the ratio of compute time to data analysed is a factor of \num{40}.
This is slow compared to the \minorchanged{\OMICRON} software \citep{Robinet:2020lbf} and represents a significant disadvantage to the hierarchical inference approach described herein.

\subsection{Level-II inference: rate only}
\label{sec:levelII}

To infer the posterior distribution on the hyperparameters \hyperparameters (i.e. \cref{eqn:hyperposterior}), we need to evaluate the likelihood in \cref{eqn:hyperlikelihood} and construct a prior on the hyperparameters.
In general, we will use a uniform prior on the rate parameter $\lambda$; for other hyperparameters, we will discuss them when introduced.

If we do not need to model the glitch population parameters, i.e., $\hyperparameters=\{\lambda\}$, then the likelihood can be rapidly calculated, taking only the evidence estimates from the Level-I inference.
\changed{I.e., since $\evidence(\data_i | \hyperparameters, \modelG) = \evidence(\data_i | \hyperparametersfixed, \modelG)$ when the population is not modelled, the per-segment glitch evidences entering \cref{eqn:hyperlikelihood} are exactly those computed in \cref{eqn:evidence}, and no re-weighting is required.}
Typically, these inferred evidences have an uncertainty, but we choose to neglect this in this work.

For the rate-only Level-II inference, the inference problem is one-dimensional.
Therefore, with a uniform prior on $\lambda$, we construct the hyperlikelihood in \BILBY and opt to evaluate the posterior on a regularly-spaced grid; this avoids sampling and is sufficiently computationally cheap to immediately produce smooth posterior distributions.

\subsection{Level-II inference: rate and population}
\label{sec:pm}
For cases where we also wish to model the glitch population parameters, each evaluation of the hyperlikelihood in \cref{eqn:hyperlikelihood} for some \hyperparameters naively requires \changed{re-evaluating the per-segment glitch evidence $\evidence(\data_i | \hyperparameters, \modelG)$, i.e.\ the integral in \cref{eqn:evidence} under $\hat{\hyperparameters}=\hyperparameters$, for all \Nseg segments}.
In general, this is computationally intractable for gravitational-wave inference problems.

Therefore, it is common to use the ``recycling'' trick introduced in \citet{Thrane:2018qnx}.
\changed{Since \cref{eqn:hyperlikelihood} requires only the per-segment evidences, it suffices to derive the recycled form for a single segment.
We begin by writing the glitch evidence for segment $i$ as the integral over the likelihood and prior:}
\begin{align}
\changed{\evidence(\data_i | \hyperparameters, \modelG)}
& = \changed{\int \likelihood(d_i | \parameters, \hyperparameters) \pi(\parameters | \hyperparameters) \, d\theta\,.}
\end{align}
For this derivation, we drop the conditional dependence on \modelG for all terms on the right-hand side.
Then, we note that the likelihood is independent of \hyperparameters, and therefore we can replace it with the likelihood under \hyperparametersfixed and expand:
\begin{align}
\changed{\evidence(\data_i | \hyperparameters, \modelG)}
& = \changed{\int \likelihood(d_i | \parameters, \hyperparametersfixed) \pi(\parameters | \hyperparameters) \, d\theta}\\
& = \changed{\int \frac{\evidence(d_i | \hyperparametersfixed) p(\parameters | \hyperparametersfixed, d_i)}{\pi(\parameters | \hyperparametersfixed)} \pi(\parameters | \hyperparameters) \, d\theta} \\
& = \changed{\evidence(d_i | \hyperparametersfixed) \int p(\parameters | d_i, \hyperparametersfixed) \frac{\pi(\parameters | \hyperparameters)} {\pi(\parameters | \hyperparametersfixed)} \, d\theta\,.}
\end{align}
We now see that, given $\Msamples_i$ posterior samples $\theta_i^\ell \sim p(\theta | d_i, \hyperparametersfixed, \modelG)$ from the individual event analysis calculated under \hyperparametersfixed, we can approximate the integral using importance sampling:
\begin{equation}
\int p(\parameters | d_i) \frac{\pi(\parameters | \hyperparameters)}{\pi(\parameters | \hyperparametersfixed)} \, d\parameters
\approx \frac{1}{\Msamples} \sum_{\ell=1}^{\Msamples} \frac{\pi(\parameters_i^\ell | \hyperparameters)}{\pi(\parameters_i^\ell)}\,,
\end{equation}
which is exact as the number of samples $\Msamples \to \infty$.

Then, the \changed{recycled glitch evidence entering the mixture in} \cref{eqn:hyperlikelihood} is computed from
\begin{align}
   \changed{\evidence(\data_i | \hyperparameters, \modelG)
   = \evidence(\data_i | \hyperparametersfixed, \modelG) \,
   \frac{1}{\Msamples_i}\sum_{\ell=1}^{\Msamples_i}
   \frac{\prior(\parameters_i^\ell | \hyperparameters)}{\prior(\parameters_i^\ell | \hyperparametersfixed)}\,,}
    \label{eqn:recycling}
\end{align}
where $\Msamples_i$ is the number of samples from the $i$th posterior distribution.
\changed{The full hyperlikelihood then follows by substituting \cref{eqn:recycling} into \cref{eqn:hyperlikelihood}.}
We note that this recycling approach is only effective when the resampled posterior for individual events is sufficiently close to the \minorchanged{Level-I} posterior calculated for $\hyperparameters=\hyperparametersfixed$.

While the recycled likelihood is faster than marginalizing over both Level-I and Level-II at the same time, this step can nevertheless introduce significant additional computational overhead.
This overhead can be reduced by using GPU parallelisation
\citep{2025JOSS...10.7753T}.
However, in this work, we introduce a new approximation method, suitable for one-dimensional population analyses or when the posteriors are uncorrelated, which we refer to as \textbf{\acl{HIQC}} (\acs{HIQC}).

\begin{figure}
    \centering
    \includegraphics[width=\linewidth]{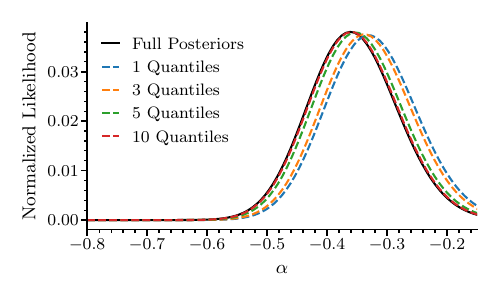}
    \caption{Validation of the \ac{HIQC} method for a toy model comparing the full likelihood, \cref{eqn:recycling} with the \ac{HIQC} likelihood, \cref{eqn:recyclying_quantiles} varying the number of quantiles.
    }
    \label{fig:quantiles_validation}
\end{figure}

Assuming that the population priors vary smoothly across the parameter space, we partition the posterior samples into $\Quantiles$ quantiles.
Let $\mathcal{Q}_k$ denote the $k$-th quantile region, defined by:
\begin{equation}
\mathcal{Q}_k = \{\theta : F^{-1}\left(\frac{k-1}{Q}\right) < \theta \leq F^{-1}\left(\frac{k}{Q}\right)\}\,,
\end{equation}
where $F$ is the empirical CDF of the posterior samples.

Within each quantile $\mathcal{Q}_k$, we approximate the integrand as constant, using a representative value $\bar{\theta}_k$ given by the median of samples in that quantile.
That is, we approximate
\begin{align}
\int_{\mathcal{Q}_k} p(d_i | \theta_i) \frac{\pi(\theta_i | \hyperparametersfixed)}{\pi(\theta_i)} \, d\theta_i
&\approx \frac{\pi(\bar{\theta}_k | \hyperparametersfixed)}{\pi(\bar{\theta}_k)} \cdot w_k\,,
\end{align}
and $w_k$ is the weight of quantile $k$
\begin{equation}
w_k = \int_{\mathcal{Q}_k} p(d_i | \theta_i) \, d\theta_i \approx \frac{\Msamples_k}{\Msamples}\,,
\end{equation}
with $\Msamples_k$ being the number of posterior samples in quantile $k$.

Combining all quantiles, we get
\begin{equation}
\int p(d_i | \theta_i) \frac{\pi(\theta_i | \hyperparametersfixed)}{\pi(\theta_i)} \, d\theta_i \approx \sum_{k=1}^{Q} w_k \frac{\pi(\bar{\theta}_k | \hyperparametersfixed)}{\pi(\bar{\theta}_k)}
\end{equation}

The HIQC hyper-log-likelihood which approximates \cref{eqn:recycling} is then given by
\begin{equation}
\log \mathcal{L}(\boldsymbol{\Lambda} | \{d_i\}) = \sum_{i=1}^{N} \log \left[ \sum_{k=1}^{\Quantiles} w_{ik} \frac{\pi(\bar{\theta}_{ik} | \boldsymbol{\Lambda})}{\pi(\bar{\theta}_{ik})} \right]\,.
\label{eqn:recyclying_quantiles}
\end{equation}
We present this in the one-dimensional case, which we use later in this work. In principle, the idea could be extended to multi-dimensional population analyses by taking a product over the likelihoods and assuming the parameters to be uncorrelated. In practice, (e.g., for modelling the \ac{BBH} mass-spin population), this will be a poor assumption.

The quantile approximation reduces computational cost from $\mathcal{O}(\Nseg\Msamples)$ to $\mathcal{O}(\Nseg \Quantiles)$ evaluations of the population distribution per likelihood call, where typically $\Quantiles \ll \Msamples$ (e.g., $Q=10$ vs $M=1000$).

To validate the method, we create a toy model in which true values are drawn from a truncated power law distribution with exponent $\alpha=1/2$.
For each true value, we create mock posteriors drawn from a bimodal distribution with a 1:10 weighting between modes.
In \cref{fig:quantiles_validation}, we plot the normalized likelihood for the full likelihood and the HIQC likelihood with different choices of $\Quantiles$.
From this, it can be seen that for $\Quantiles=1$, there is a mild bias in the likelihood, but with $\Quantiles=10$, HIQC accurately recovers the full likelihood.
While this example does not validate the method in full generality, it demonstrates HIQC can accurately approximate the full likelihood and produce a speed-up of $\Msamples/\Quantiles$.

For the rate and population Level-II inference, we use \BILBY to sample the posterior distribution on the hyperparameters, using the \DYNESTY nested sampling algorithm.

\subsection{Level-II inference: time-dependent}
\label{sec:td}
Over periods longer than a few tens of minutes, we do not expect that the glitch rate $\lambda$ is constant for interferometric detectors \changed{(this is supported by the results we present later in \cref{sec:results})}.
Therefore, we extend our Level-II inference to model a time-varying glitch rate, $\lambda = \lambda(t)$.
There are numerous ways this can be done, but we choose a parametric basis function representation:
\begin{align}
    \lambda(t) = \lambda_0 10^{f(t; \basisparameters)}\,,
\end{align}
where $f(t)$ is a sum of suitable basis functions with parameters \basisparameters to model the temporal variations. The baseline glitch rate $\lambda_0$ and basis parameters \basisparameters will be inferred during the Level-II inference process.

\changed{This modifies the hyperlikelihood of \cref{eqn:hyperlikelihood} in one respect.
With a constant rate, the glitch probability $P(\modelG | \hyperparameters)$ is common to every segment; once $\lambda = \lambda(t)$, it becomes segment-dependent, and the mixture weight for segment $i$ is evaluated at the central time $t_i$ of that segment:
\begin{align}
    P_i(\modelG | \hyperparameters) = \lambda(t_i) \delta t \exp\left(-\lambda(t_i) \delta t\right)\,.
    \label{eqn:pglitch_td}
\end{align}
Because the mixture in \cref{eqn:hyperlikelihood} is applied segment by segment, this substitution requires no further change to the structure of the hyperlikelihood: the per-segment evidences $\evidence(\data_i | \hyperparameters, \modelG)$ and $\evidence(\data_i | \modelN)$ are those already computed in the Level-I analysis, and only the weights with which they are combined acquire a time dependence.}

One simple basis is a truncated Fourier series:
\begin{align}
   f(t) = \sum_{j=1}^{\Nbasis} c_j \sin\left(2\pi j \tilde{t}\right)\,,
\end{align}
where $\tilde{t}$ is the min-max normalization of the time $t$:
\begin{align}
    \tilde{t} = \frac{t - \mathrm{min}(t)}{\mathrm{max}(t) - \mathrm{min}(t)}\,.
\end{align}
This basis set contains $\Nbasis$ additional coefficients, the amplitudes of the harmonic sinusoids.

A slightly more complicated, but expressive basis, is the B-spline of order $k$:
\begin{align}
    f(t) = \sum_{j=1}^{\Nbasis} c_j B_{j, k}(t)\,
\end{align}
where the knots of the spline are distributed uniformly over the observation duration.

As with the rate-only and rate+population Level-II inference, we use \BILBY to sample the posterior distribution on the hyperparameters, which now include the time-dependent rate parameters.
The time-dependent rate can also be coupled with the population analysis to perform a joint inference of the population properties and time-dependent rate.
In principle, the population properties could also be modelled similarly, providing a time-dependent distribution.
However, in this work, we restrict ourselves to a time-dependent rate and a time-independent population distribution.

Once the basis function parameters are inferred, we can reconstruct the time-dependent rate $\lambda(t)$ and its uncertainty by sampling from the posterior distribution of the basis parameters.
From this, we can create a plot showing the inferred rate as a function of time, with uncertainty bands.

\begin{figure*}
    \centering
    \includegraphics[width=\linewidth]{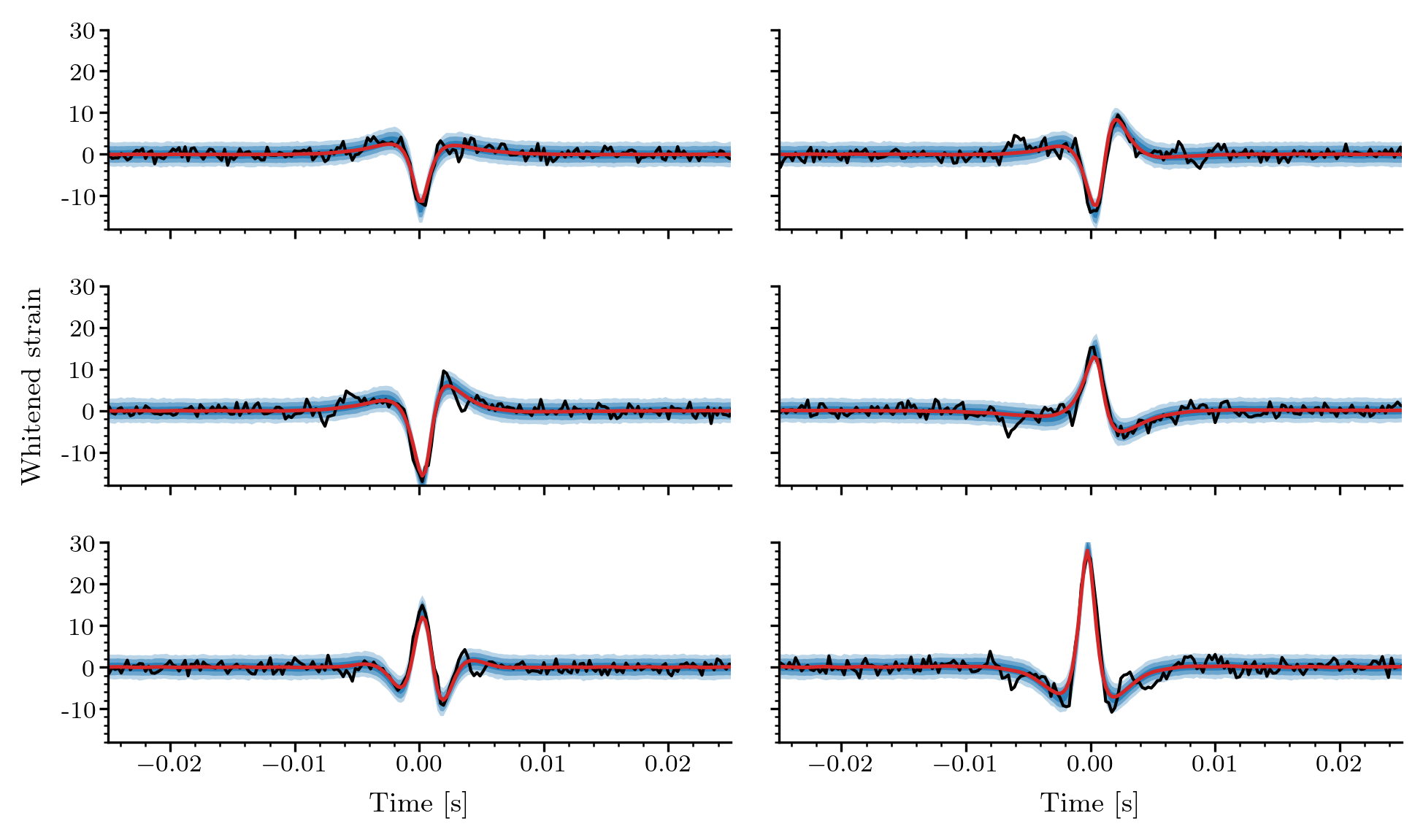}
    \caption{The whitened strain and posterior predictive distribution from six realisations of the simulated data used in our validation study.
    In black, we show the data, after whitening, which includes coloured Gaussian noise alongside a glitch simulated by \GLITCHFLOW.
    On top, we add the prediction of the \ANTIGLITCH model under the maximum posterior estimate (red curve) and shaded blue bands showing the 68, 95, and 99\% uncertainty levels.
    }
    \label{fig:demonstration}
\end{figure*}

\section{Validation studies}
\label{sec:validation}

We perform a series of validation studies to investigate the performance of our algorithm under controlled conditions.
To simulate the continuous background Gaussian noise present in the detectors, we employ the standard approach: colouring white Gaussian noise in the frequency domain and then Fourier transforming to the time domain.
However, we point out that this choice does not capture the non-stationary nature of real detector noise, nor does it fully model the coherent lines (e.g., from the power supply and violin modes).
For this validation study, we use \BILBY to generate the noise, using the \ac{LHO} \ac{PSD} from the first three months of the third observing run \citep{noise_curves}.

To simulate glitches, we initially perform a study adding glitches modelled by the \ANTIGLITCH model to coloured Gaussian noise.
We then run the \minorchanged{Level-I} inference on these simulated data segments and check that the observed posterior distributions were consistent with the simulated parameters, confirming the correct implementation of our algorithm.
We do not present the results here as the validation step itself is subsumed by later validation studies.

To add realism to our validation study, we simulate glitches by drawing them from a normalising flow \citep{Kobyzev:2019ydm}, trained on Blip glitches \citep{Cabero:2019orq} from the third observing run.
The initial implementation of this approach was described in \citet{Malz:2025xdg} and used to perform joint astrophysical signal and glitch inference: we extract the glitch modelling tool from the framework, named \GLITCHFLOW, developed in that work, and use it here for glitch simulation. 
Using the \GLITCHFLOW framework with the same settings as described in \citet{Malz:2025xdg}, we train a new model on O3 Blip glitches from the \GRAVITYSPY catalogue \citep{Zevin:2016qwy,2024EPJP..139..100Z}. The training is done on 1/8 s of whitened time series data surrounding each glitch. Singular value decomposition is used to factorise the training data into a basis matrix and associated weights, and the normalising flow is trained on the weights only. Once trained, we sample from the flow and reconstruct the glitch using the basis matrix, thus creating a set of glitches representative of the training data. 
By using \GLITCHFLOW in this way, we capture the salient features and variety of the glitches in the class, without needing to build a population model.
There is a single tunable parameter that we introduce: a scaling parameter, which we use to control the amplitude of the simulated glitches.
This approach also allows us to precisely control the distribution of glitches simulated in a continuous stretch of data in a way that is not possible when using real data.

\changed{Blips are one of several glitch classes present in the data, and our validation therefore establishes the performance of the method for short-duration, blip-like morphologies, which is precisely the regime in which \ANTIGLITCH is expected to be a good model.
It should not be read as a validation against low-frequency or scattering-like glitches.
For an indication of how the method behaves when the morphology falls outside this class, see the high-\ac{SNR} triggers recovered with $\lnB<0$ in \cref{app:examples}.}

\subsection{Single glitch analysis}
\label{sec:validation_sga}

To demonstrate the performance of our Level-I analysis using the \ANTIGLITCH model on glitches simulated from \GLITCHFLOW, in \cref{fig:demonstration} we show six random realisations of simulated data containing Gaussian noise and glitches drawn from \GLITCHFLOW. The resulting strain is whitened, and we include fits resulting from the Level-I analysis.
We find that the fits capture the broad features of the simulated glitches, in agreement with \citet{Bondarescu:2023jcx}, which found the model to be well-suited to model blip glitches.
This plot also demonstrates the correct implementation of the time and phase marginalisation and reconstruction, since the predicted models are generated from the reconstructed posterior distributions.

\subsection{Hierarchical Bayesian model validation}
\label{sec:validation_hierarchical}
\begin{figure}
    \centering
    \includegraphics[width=\linewidth]{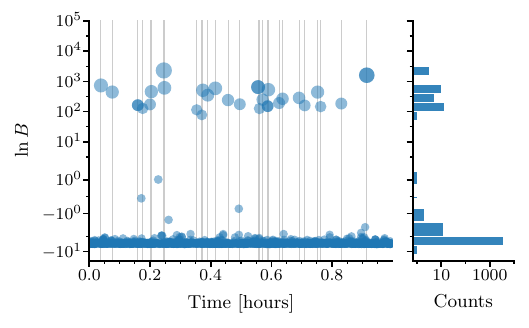}
    \caption{The distribution of Log Bayes factors (\lnB, i.e., the ratio of evidence for the \ANTIGLITCH model against the coloured-Gaussian noise only hypothesis) calculated for data containing coloured Gaussian noise and glitches simulated with \GLITCHFLOW.
    In the left-hand figure, we show the distribution of \lnB as a function of time, with the marker size scaled by \lnB and vertical lines marking the times of simulated glitches.
    In the right-hand figure, we show a histogram of the \lnB values.
    }
    \label{fig:constant_glitch_rate_data}
\end{figure}

To validate the hierarchical inference method, we generate \qty{4096}{\second} of data and simulate glitches using \GLITCHFLOW in this data following a Poisson process with a constant rate of \qty{0.01}{\Hz}.
We then analyse \qty{1}{hour} of this data (offset from the start to enable \ac{PSD} estimation) with the Level-I inference approach.
In \cref{fig:constant_glitch_rate_data}, we plot the Bayes factors for each Level-I analysis as a function of the analysis time and as a histogram.
This can be compared with the times of simulated glitches, and we find that the Bayes factors are large for all simulated glitches: the distribution of Bayes factors is bimodal, with a clear separation between the noise-only segments and the glitch segments.

\begin{figure}
    \centering
    \includegraphics[width=\linewidth]{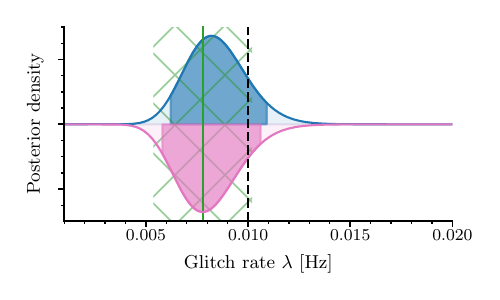}
    \caption{Estimates of the glitch rate calculated over the \qty{1}{hour} of data presented in \cref{fig:constant_glitch_rate_data}.
    A violin plot is used: the upper half is the posterior under the hierarchical \minorchanged{Level-II} analysis described in \cref{sec:levelII} while the lower half is the posterior under the Gamma distribution described in \cref{sec:counting}, but applied to the set of simulated glitch times.
    For both posterior distributions, the darker shaded region marks the symmetric 90\% uncertainty region (calculated from the 0.05-0.95 quantile range).
    We also show the mean estimate from the simple counting rate predicted by \cref{eqn:lambda_hat} and applied to the simulated glitch times (solid vertical green line), and the $90\%$ uncertainty calculated from scaling the uncertainty estimate in \cref{eqn:lambda_hat_err} (hatched green lines).
    Finally, a vertical dashed line marks the true glitch rate used in the simulation.
    }
    \label{fig:constant_glitch_rate_violins}
\end{figure}

Next, in \cref{fig:constant_glitch_rate_violins}, we take the \qty{1}{\hour} of data analysed in \cref{fig:constant_glitch_rate_data} and infer the posterior distribution of the glitch rate using the rate-only Level-II analysis described in \cref{sec:levelII}.
The 90\% credible interval on the glitch rate contains the true rate, indicating the method has been successful.

Next, we take the times of simulated glitches in the data, and apply the two counting-based approaches described in \cref{sec:counting}, plotting the resulting Gamma conjugate posterior and the credible interval prediction from a simple Poisson rate estimation.
We refer to these two estimates as the \emph{measurable} rate to indicate they are the rate one could infer with perfect knowledge of the true number of glitches (but not the true rate).

These three analyses show good agreement, though there are modest differences: this is expected since all three have different underlying assumptions and differing levels of information about the glitch population.
This figure demonstrates that the hierarchical Bayesian method can analyse the full data set (with no threshold) and infer the rate with equivalent accuracy to that obtained from knowing the actual times of the simulated signals.
However, this result is less remarkable since, as seen in \cref{fig:constant_glitch_rate_data}, the Bayes factors clearly separate the glitches from the background noise, and there are many tens of glitches. So, even a simple counting rate analysis would also be effective since a simple threshold could be applied to the Bayes factors to separate the glitches from the noise.

To investigate the capacity of the hierarchical model to infer the glitch rate when the number of observed glitches is reduced, we recycle the data from \cref{fig:constant_glitch_rate_data}, taking subsets of different durations.
We then compare the posterior from the full hierarchical model, again with the measurable glitch rate inferred using the Gamma conjugate posterior and the simple Poisson counting method.
This demonstrates that, when the duration is greater than ten minutes or so (for which $\lambda T \gg 1$), the measured rate agrees with the measurable rate using either the Poisson or Gamma methods.
However, for short durations, where there are few events, the simple Poisson counting method fails as expected, with the posterior on the glitch rate incorrectly concentrating near zero, despite the paucity of data.
On the other hand, both the measurable rate inferred from the Gamma conjugate posterior and the measured rate from the hierarchical Bayesian model closely agree and tend to large values of $\lambda$, reflecting the true uncertainty in the rate as $T\rightarrow 0$.

\begin{figure}
    \centering
    \includegraphics[width=\linewidth]{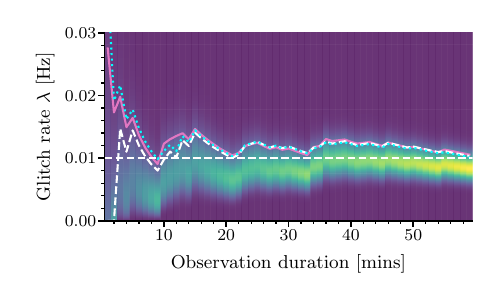}
    \caption{The inferred glitch rate for a fixed rate, but varying observation duration.
    For each vertical slice at duration $T$, we take the first $T$ minutes of data from \cref{fig:constant_glitch_rate_data} and then infer the posterior distribution on the rate; this is shown as a color map. We additionally add curves showing the 90\% upper limit on the rate inferred from the hierarchical model (pink), Gamma conjugate posterior (cyan), and simple Poisson counting method (white).
    Finally, a horizontal dashed line marks the true glitch rate used in the simulations. 
    }
    \label{fig:constant_glitch_rate_heatmap}
\end{figure}

\begin{figure}
    \centering
    \includegraphics[width=\linewidth]{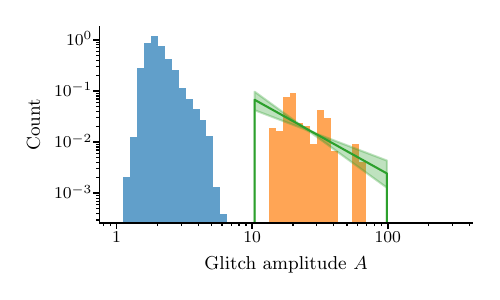}
    \caption{Inferred amplitude distributions for noise-only (blue) and glitch (orange) simulations, along with the predicted power-law scaling (green).}
    \label{fig:constant_glitch_rate_population_amplitude}
\end{figure}

For a final analysis of the data presented in \cref{fig:constant_glitch_rate_data}, we utilise the \ac{HIQC} population inference method described in \cref{sec:pm} to measure the rate and population properties of the glitches.
This generalises the analysis in \cref{fig:constant_glitch_rate_violins}, allowing modelling of the population properties of the glitches.
We model the amplitude distribution of the population with a power-law, having a single variable, the exponent $\alpha$ (we keep the minimum and maximum of the distribution fixed at \num{10} and \num{100}).
Analysing the data, we find the two-dimensional posterior distribution to be broadly uncorrelated with the inferred amplitude exponent $\alpha=-1.47^{+0.26}_{-0.27}$.
In \cref{fig:constant_glitch_rate_population_amplitude}, we create a predictive distribution plot showing the median inferred amplitude of noise-only simulations and glitch simulations.
On this figure, we then plot the inferred power-law scaling of the amplitude, finding it to be broadly in agreement with the glitch distribution.
Moreover, in \cref{fig:constant_glitch_rate_population_amplitude}, we also note that the amplitude distributions of the noise and glitch simulations are easily separated.
This is by design from our choice of scaling parameter.
While it does not reflect the expected distribution of glitches in real data, it allows us to validate the method in a regime where the glitch and noise distributions are cleanly separated.

\begin{figure}
    \centering
    \includegraphics[width=\linewidth]{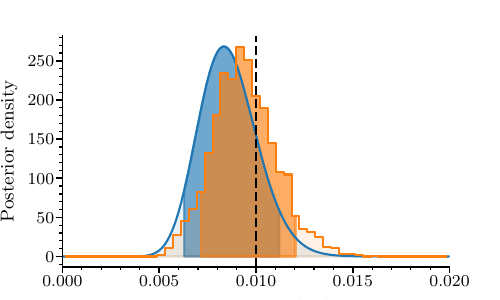}
    \caption{Posterior distributions inferred from the \minorchanged{Level-II} analysis of the data presented in \cref{fig:constant_glitch_rate_data} showing the rate-only inference (blue) and rate and population inference (orange), in which the amplitude distribution is also modelled.
    Darker shaded regions demark the 90\% credible interval for both distributions, while a vertical dashed line marks the simulated glitch rate.
    }
    \label{fig:constant_glitch_rate_population_comparison}
\end{figure}

Finally, in \cref{fig:constant_glitch_rate_population_comparison}, we then plot the posterior on the inferred rate using the rate and population \minorchanged{Level-II analysis} and compare it to the rate-only analysis.
When modelling the glitch population, we find a slightly larger glitch rate, improving the consistency with the true value (though this could be due to statistical fluctuations in the simulated data).

\subsection{Sensitivity analysis}
\label{sec:validation_sensitivity}
The key advantage of the hierarchical inference approach presented in this work is that it does not require a user-specified \ac{SNR} threshold.
As a result, it can use the entire data stream and, we claim, therefore retain measurement accuracy for weaker signal distributions when compared to a counting-based alternative.
To test this claim, we now design a sensitivity study.

We begin by creating a set of simulated data to use as a source for our sensitivity study.
The simulation data is created using the method presented in \cref{sec:validation_hierarchical}, but we vary the distribution of the simulated glitch amplitudes so that the signal and noise distributions overlap.
(This was not the case in \cref{fig:constant_glitch_rate_data}, where we can see that the scaling parameter is sufficiently large that the \lnB values cleanly separate the glitches from the Gaussian noise background).
This is done by randomly drawing the scaling parameter for each simulated glitch from a range of values that we empirically find to cover the regime between small to moderate \ac{SNR}.

\begin{figure}
    \centering
    \includegraphics[width=\linewidth]{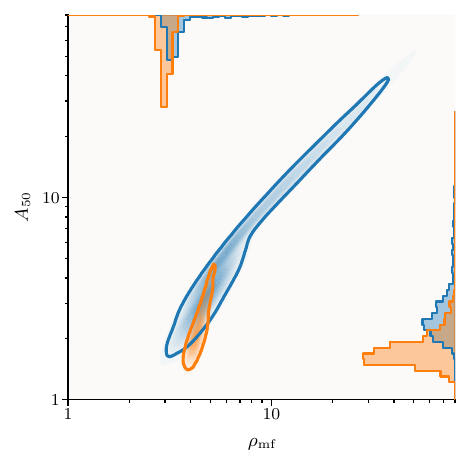}
    \caption{Distribution of the inferred maximum \rhomf and median amplitudes measured from the Level-I analysis of the data simulated for the sensitivity study.
    We separate the distributions into the inferences from data segments including a glitch (blue) and data segments with only Gaussian noise (orange).}
    \label{fig:sensitivity_rho_mf_A50_distribution}
\end{figure}

For each simulated data set, we apply the \minorchanged{Level-I} analysis.
In \cref{fig:sensitivity_rho_mf_A50_distribution}, we provide histograms showing the one and two-dimensional density of the measured matched-filter \acp{SNR} and amplitudes, separated by noise-only segments and segments that contain glitches.
This figure validates our range of scaling parameters, showing that our ``glitch'' distribution overlaps with the fixed ``noise'' distribution in both the inferred amplitudes and \rhomf.
It also demonstrates the expected scaling between amplitude and \ac{SNR}.

Next, we take subsets of results from the \minorchanged{Level-I} analyses presented in \cref{fig:sensitivity_rho_mf_A50_distribution}, but varying, by hand, the statistical properties of the glitch distribution.
For all subsets, the effective duration is fixed at \qty{1000}{\second} and the true glitch rate is fixed to be $\lambda=$\qty{0.1}{\hertz} such that the expected number of glitches is $100$.
However, for each subset, we filter the set of glitches, applying a minimum to the matched-filter \ac{SNR} \rhomf of the signal distribution, before sampling the subset.
In effect, this modifies the glitch distribution from which we sample.
This allows us to investigate how the inferred rate varies with the minimum \rhomf of the signal distribution.

\begin{figure}
    \centering
    \includegraphics[width=\linewidth]{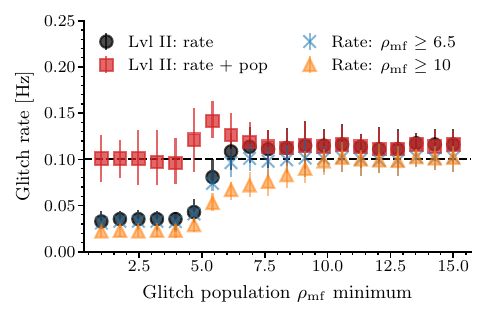}
    \caption{The inferred glitch rate taken from a \qty{1000}{\second} set of data randomly sampling from the signal and noise distributions in \cref{fig:sensitivity_rho_mf_A50_distribution}, but varying the minimum \rhomf of the signal distribution.
    We compare the rate inferred from the \minorchanged{Level-II} analysis (black) with the rate inferred from a Poisson counting rate using thresholds of 5 (blue), 6.5 (orange), and 10 (green).
    For each choice of minimum \rhomf, we repeat the study 20 times.
    Circles denote the median of the median over the repeated studies, while the error bars denote the 90\% interval of the median over the repeated studies.
    Finally, for the \minorchanged{Level-II} analysis, we also include a shaded band denoting the median of the 90\% interval from the underlying distribution, averaged over the repeated studies.
    }
    \label{fig:sensitivity_rho_mf_rate}
\end{figure}

Finally, taking each subset of data, we then infer the rate using three methods and plot the results in \cref{fig:sensitivity_rho_mf_rate}.

First, we use the simple Poisson rate, taking thresholds of \num{6.5} and \num{10}.
For these two inferred rates, when the minimum \rhomf is greater than the threshold, they accurately recover the correct glitch rate.
This is to be expected since the noise distribution has a steep right-hand tail in the \ac{SNR} with no noise-only simulations having $\rhomf \gtrsim 6$ (cf. \cref{fig:sensitivity_rho_mf_A50_distribution}).
As a result, the noise and glitch distributions are cleanly separated, and the simple Poisson counting rate can easily distinguish the two different distributions.
However, for this counting-rate approach, once the minimum \rhomf of the glitch distribution falls below the threshold (either \num{6.5} or \num{10}), the rate becomes biased with an increasing number of false negatives leading to an underestimate of the true rate.

Second, we apply the rate-only Level-II analysis (cf. \cref{sec:levelII}).
The behaviour of this method in \cref{fig:sensitivity_rho_mf_rate} closely matches that of the Poisson rate with a threshold of \num{6.5}:
The inferred rate is accurate above a minimum \rhomf of $\sim 6$, but biased toward an underestimate of the rate below.
The cause of this can be understood from \cref{fig:sensitivity_rho_mf_A50_distribution}: this threshold approximately corresponds to the point where the noise and glitch distributions begin to overlap (as the minimum \rhomf is decreased).
We diagnose that the cause of the bias is mismodelling: in this rate-only analysis, the prior used in the Level-I analysis (uniform in amplitude) determines the assumed population properties of glitches.
However, we highlight that, unlike the Poisson rate, no threshold was pre-determined for this rate-only analysis: the method naturally finds a bias at this point due to mismodelling.

Finally, we correct for this mismodelling by applying a rate and population Level-II analysis as described in \cref{sec:pm}.
To model the amplitude distribution, we use a power-law distribution which has three free parameters: the minimum, maximum, and exponent.
To reduce the computation time, we forego a full analysis, marginalising over all three parameters, and instead use fixed values.
For the minimum and maximum, we take rounded values of the inferred distribution from the simulation data set.
Meanwhile, for the exponent, we choose a fixed value of \num{-1.6} which we optimise by visual inspection of \cref{fig:sensitivity_rho_mf_rate}.
This is a sub-optimal approach that cannot be reproduced on real data.
However, what we see in \cref{fig:sensitivity_rho_mf_rate} the rate and population method remains unbiased when the distributions substantially overlap.
There is a region around \minorchanged{$\rhomf \sim 5$} where this is not true: we expect that this is due to our sub-optimal approach.
Fully marginalizing over the population model will be a requirement for the analysis of observational data.

In summary, our sensitivity study demonstrates that the full hierarchical Bayesian model can measure the rate even when the glitch and noise distributions overlap.
However, it is vital to measure the population properties to do so.
Without a population model, mismodelling of the glitch distribution can bias the inferred rate.
Nevertheless, this still may be useful as it provides a means to infer the rate without a threshold, subject to assumptions on the amplitude distribution.

\subsection{Time-dependent rate analysis}
\label{sec:val_td}

\begin{figure}
    \centering
    \includegraphics[width=\linewidth]{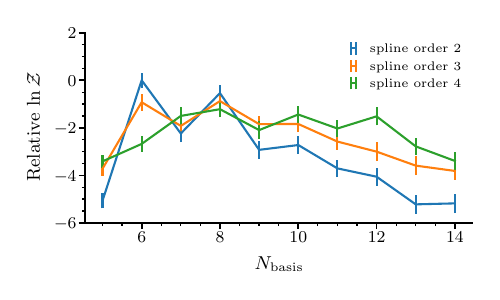}
    \caption{Natural-log-evidence as a function of the number of bases used in the \minorchanged{Level-II} analyses applied to simulated data in which the glitch rate suddenly changes.
    All values are given relative to the maximum, which is found at $N_{\rm basis}=6$ with a spline order of \num{2}.}
    \label{fig:variable_glitch_rate_nbasis}
\end{figure}

To validate the time-dependent \minorchanged{Level-II} inference method described in \cref{sec:td}, we follow the data-generating procedure described in \cref{sec:validation_hierarchical}, but now generate \qty{8192}{\second} of data with a step change in the glitch rate from \qty{0.01}{\hertz} to \qty{0.05}{\hertz}.
We then apply the \minorchanged{Level-I} analysis to a \qty{2}{\hour} stretch of the simulated data containing the sudden change at approximately the midway point.
Next, we run the \minorchanged{Level-II} analysis using spline bases varying the spline order and number of bases.
In \cref{fig:variable_glitch_rate_nbasis}, we plot the natural log evidence estimates for each analysis relative to the maximum found across the set of analyses.
In effect, this enables a Bayes factor comparison which demonstrates that 6 basis functions with a spline order of \num{2} is the optimal model.
By design, such an evidence maximisation accounts for model complexity by virtue of the implicit Occam factor weighting against more complicated models (this is visible in \cref{fig:variable_glitch_rate_nbasis} as the increasing number of bases decreases the relative evidence).

\begin{figure*}
    \centering
    \includegraphics[width=\linewidth]{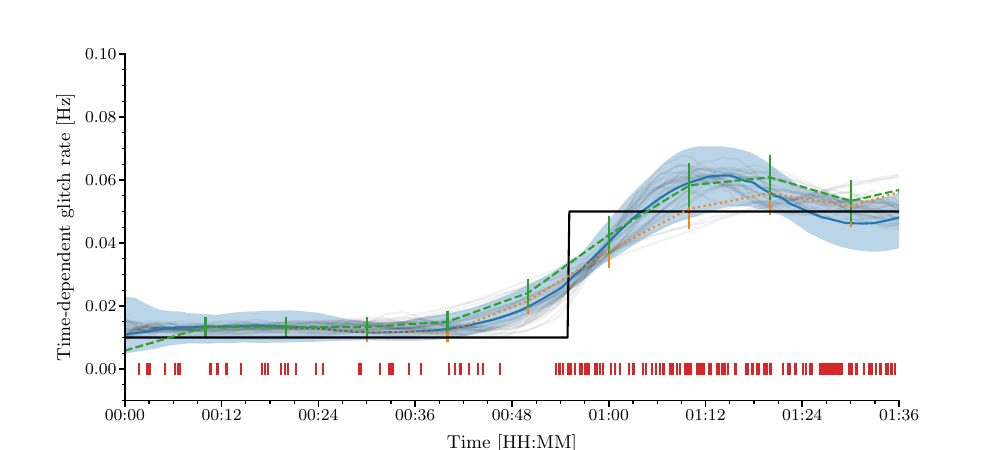}
    \caption{The time-dependent glitch rate inferred from simulated data in which the glitch rate suddenly changes.
    A solid black line marks the true glitch rate while red vertical bars denote the time of simulated glitches.
    The blue solid line marks the median rate inferred from the time-dependent \minorchanged{Level-II} analysis using splines with $N_{\rm basis}=8$; a blue band marks the 90\% interval on the inferred distribution.
    Similarly, the orange line and band mark the rate inferred from a time-dependent \minorchanged{Level-II} analysis combined with a population analysis.
    A dotted red line marks the measurable rate inferred using a Poisson counting rate estimation from the set of simulated glitch times (using \qty{1200}{\second} bins) with error bars marking the 68\% interval.
    }
    \label{fig:variable_glitch_rate_lambdat}
\end{figure*}

In \cref{fig:variable_glitch_rate_lambdat}, we show the predicted glitch rate inferred from the optimal model, along with the measurable rate inferred from the set of simulated glitch times.
We also add the median estimate from all models computed in \cref{fig:variable_glitch_rate_nbasis}, which demonstrates that the general trend is similar.
Therefore, we find that maximising the evidence is not strictly required since our primary interest is the time-dependent behaviour.

The rate measured using the time-dependent \minorchanged{Level-II} analysis clearly identifies the step change.
However, the spline bases are unable to infer the exact sudden step change and smooth out the behaviour.
This is to be expected since the splines smoothly interpolate an underlying continuous function.
The \minorchanged{Level-II} analysis closely agrees with the binned estimate of the measurable rate, which is also smoothed by the sliding window.

\section{Results: one day of data from LIGO Livingston}
\label{sec:results}
To study the performance of our method in practice, we select one \qty{24}{\hour} period of data from \ac{LLO} during the O4 observing run \citep{Capote:2024rmo, LIGO:2024kkz} and perform an analysis using our hierarchical inference method.
We select the day 2023-08-18 arbitrarily, except that the detector was in observing mode for the entire day.
We begin with a discussion of the \minorchanged{Level-I} analysis, which is performed on each second of data, and then move to the \minorchanged{Level-II} analysis, which combines the results from the \minorchanged{Level-I} analysis to infer the glitch rate and population properties.

\subsection{Level-I analysis}

\begin{figure}
    \centering
    \includegraphics[width=\linewidth]{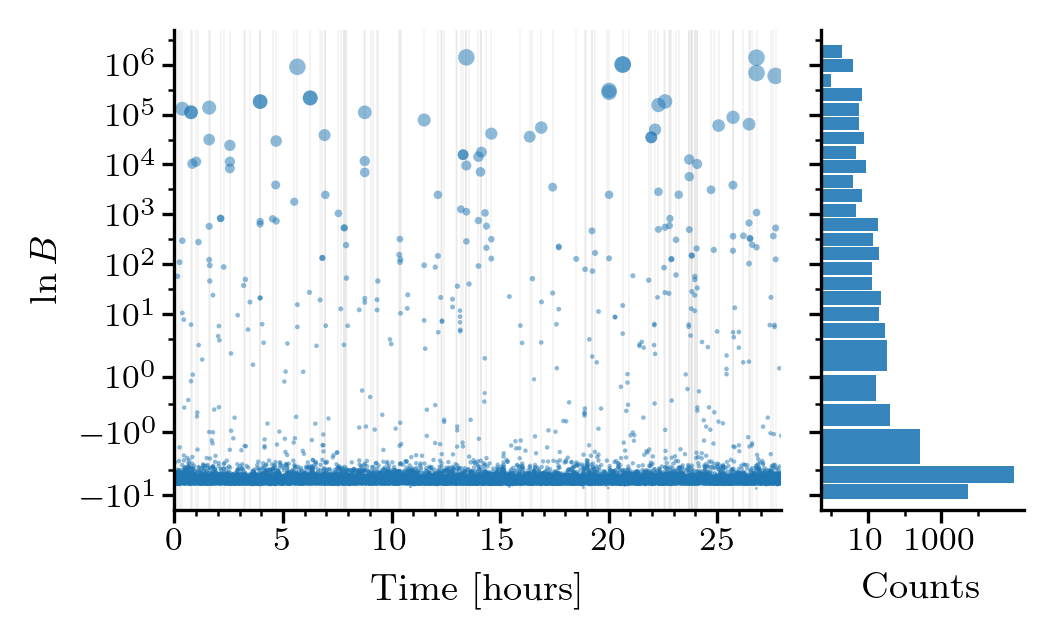}
    \caption{The distribution on \lnB (i.e. the ratio of evidence for the \ANTIGLITCH model against the coloured-Gaussian noise only hypothesis) as a function of time for one day of data from \ac{LLO} (2023-08-18), the same day as presented in \cref{fig:glitches}.
    In the left-hand figure, we show the distribution of \lnB as a function of time, with the marker size scaled by \lnB, with vertical lines denoting the times of \minorchanged{\OMICRON} triggers with \ac{SNR} greater than 10.
    In the right-hand figure, we show a histogram of the \lnB values.
    }
    \label{fig:LLO_data}
\end{figure}

We apply the \minorchanged{Level-I} analysis described in \cref{sec:levelI} to the day of data obtained from the \ac{GWOSC} O4 data release \citep{LIGOScientific:2025snk}.
In \cref{fig:LLO_data}, we visualise the measured Bayes factors during the data span and compare this with the recorded times of \minorchanged{\OMICRON} triggers (recalling that \minorchanged{\OMICRON} triggers are only recorded for \ac{SNR}~$\ge 5$).
This figure demonstrates that for the majority of \minorchanged{\OMICRON} triggers, we find a corresponding increase in the Bayes factor, indicating that the \minorchanged{\ANTIGLITCH} model consistently finds evidence for a glitch in the data.
Meanwhile, we can also visually identify a handful of cases where the Bayes factor is large, but no \minorchanged{\OMICRON} trigger is recorded.
Finally, unlike \cref{fig:constant_glitch_rate_data}, the distribution of Bayes factors is not bimodal, but instead has a long tail to large values with no distinct separation between the noise-only and glitch segments.

\begin{figure}
    \centering
    \includegraphics[width=\linewidth]{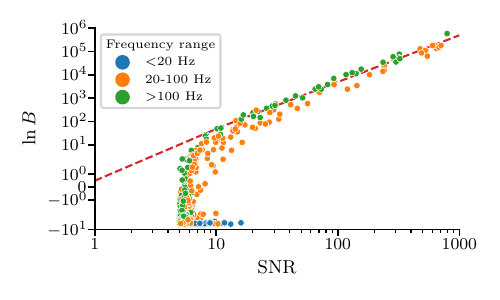}
    \caption{A comparison of the \ac{SNR} taken from \minorchanged{\OMICRON} triggers presented in \cref{fig:glitches} and the \lnB presented in \cref{fig:LLO_data}.
    Matches are found by identifying the highest-SNR \minorchanged{\OMICRON} trigger within a \qty{1.1}{\second} window of each \minorchanged{Level-I} analysis; if no \minorchanged{\OMICRON} trigger is found, the data point is not recorded in this figure. 
    An orange dashed line marks the prediction from \cref{eqn:lnBSNR} of the relationship between the Bayes factor and \ac{SNR} in the high-SNR domain.
    We note that \minorchanged{\OMICRON} triggers for \ac{SNR}$<5$ are not recorded.
    }
    \label{fig:LLO_matches}
\end{figure}

To further probe cross-matches between the \minorchanged{Level-I} analysis and the set of \minorchanged{\OMICRON} triggers with \ac{SNR}~$\ge 5$, we perform a more detailed comparison.
First, we take every \minorchanged{\OMICRON} trigger and identify the \minorchanged{Level-I} analysis that is within \qty{1.1}{\second} and then take the maximum \lnB.
In \cref{fig:LLO_matches}, we compare the SNR of the \minorchanged{\OMICRON} trigger with \lnB from the corresponding \minorchanged{Level-I} analysis.
This reveals that for high-\ac{SNR} triggers, the \ac{SNR} and \lnB are linearly correlated (in log-space).
This is expected since, in the high-\ac{SNR} domain
\begin{align}
    \ln B \approx \frac{1}{2}\rho^2\,,
    \label{eqn:lnBSNR}
\end{align}
when the glitch model is a good match to the data.
We validate this by adding the prediction to \cref{fig:LLO_matches}, showing good agreement for \ac{SNR}~$\gtrsim 20$.

However, we note that in \cref{fig:LLO_matches}, there is a population of \minorchanged{\OMICRON} triggers with \ac{SNR}~$\gtrsim 10$ that have $\lnB < 0$, indicating that the \minorchanged{\ANTIGLITCH} model is a poor fit to the data.
In Appendix~\ref{app:examples}, we show the \num{9} examples of cases where $\ln B < 0$ and \ac{SNR}~$\ge 10$, and we find that these appear to arise from low-frequency transient noise that is not well captured by the glitch model, which is designed to capture short-duration glitches.
This highlights the importance of using a glitch model that is flexible enough to capture the variety of glitches present in the data, and we discuss this further in \cref{sec:discussion}.

Finally, we take every \minorchanged{Level-I} analysis with $\ln B > 10$ and identify the \minorchanged{\OMICRON} trigger with the highest SNR within \qty{1}{\second}.
In total, we find \num{166} \minorchanged{Level-I} analyses with $\ln B > 10$, of which \num{88} have a corresponding \minorchanged{\OMICRON} trigger within \qty{1.1}{\second} (these are shown in \cref{fig:LLO_matches}).
This leaves \num{78} \minorchanged{Level-I} analyses with $\ln B > 10$ that do not have a corresponding \minorchanged{\OMICRON} trigger.

From these diagnostic figures, we can conclude that, while the \minorchanged{\ANTIGLITCH} model has failed to capture a subset of the glitches found by \OMICRON, it nevertheless finds the bulk of them and identifies a subset of times where the Bayes factor favours a glitch while \minorchanged{\OMICRON} recorded no trigger with \ac{SNR} greater than \num{5}.

\subsection{Level-II analysis: time-independent}

\begin{figure}
    \centering
    \includegraphics[width=\linewidth]{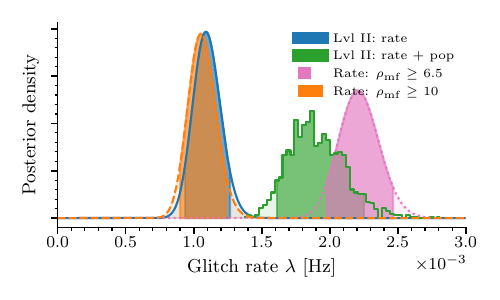}
    \caption{The glitch rate inferred from the data presented in \cref{fig:LLO_data} under the rate-only analysis (blue), rate and population (green), and the Gamma posterior from the counting rate with threshold of \num{6.5} (pink) and \num{10} (orange).
    }
    \label{fig:LLO_rate_and_pop}
\end{figure}

Taking the full day of Level-I analyses, we infer the glitch rate from the entire data set, assuming no time dependence. 
In \cref{fig:LLO_rate_and_pop}, we show the posterior distributions for the rate-only Level-II analysis, the rate and population analysis, and the rate inferred from the Poisson counting method on the \OMICRON triggers under the two standard thresholds. Given the large number of glitches in this period, the Poisson counting method produces equivalent results to the Gamma posterior method, and we therefore show only the former for clarity.

Considering first the rate-only analysis, this peaks at approximately \qty{1}{\hertz} and is similar to the rate inferred from the \OMICRON triggers using an \ac{SNR} threshold of \num{10}.
Meanwhile, the rate inferred from the \OMICRON triggers with an \ac{SNR} threshold of \num{6.5} is several times larger.

For the rate and population inference, we model the amplitude distribution as a power law with a uniform prior on the minimum between \num{1} and \num{10} and a uniform prior on the scaling exponent between \num{-2} and \num{0}.
The non-zero bound on the amplitude is required, as a bound of zero results in a posterior that favours a glitch rate several orders of magnitude larger than expected and a steep scaling exponent: in effect, the Gaussian noise distribution is interpreted as glitches.
By placing a finite lower bound, we focus the analysis only on weak, but still finite glitches.
From this analysis, we infer the minimum amplitude to be ${3.81}_{-0.89}^{+0.58}$ and the scaling exponent to be ${-1.51}_{-0.11}^{+0.11}$ (with 90\% credible intervals).
The rate is shown in \cref{fig:LLO_rate_and_pop} and demonstrates that by accounting for the population properties, we infer a larger glitch rate.
The posterior overlaps with, but is slightly below the \OMICRON rate for a threshold of \num{6.5}.

\subsection{Level-II analysis: time-dependent}
\label{sec:td_results}

For the final part of this analysis, we then analyse the day of data using the time-dependent Level-II analysis introduced in \cref{sec:td}.
Again, we use a spline basis, but based on the findings of \cref{sec:val_td}, we choose not to optimise the spline order and number of bases but instead take a fixed choice of order \num{3} and \num{24} bases (the increased numbers are chosen as the duration of data is longer).
We note that this may over-fit the data, but initially, we are interested only in the broad behaviour of the method.
In addition, we found in testing that nested sampling became slow when analysing the full data set with the time-dependent rate model coupled with population inference.
Therefore, we opt instead to use a Laplace approximation, using an optimiser to find the peak of the posterior and then sampling from a multivariate Gaussian. We find this works well and gives similar results to within the model uncertainties while providing a dramatic speedup in compute time.

\begin{figure*}
    \centering
    \includegraphics[width=\linewidth]{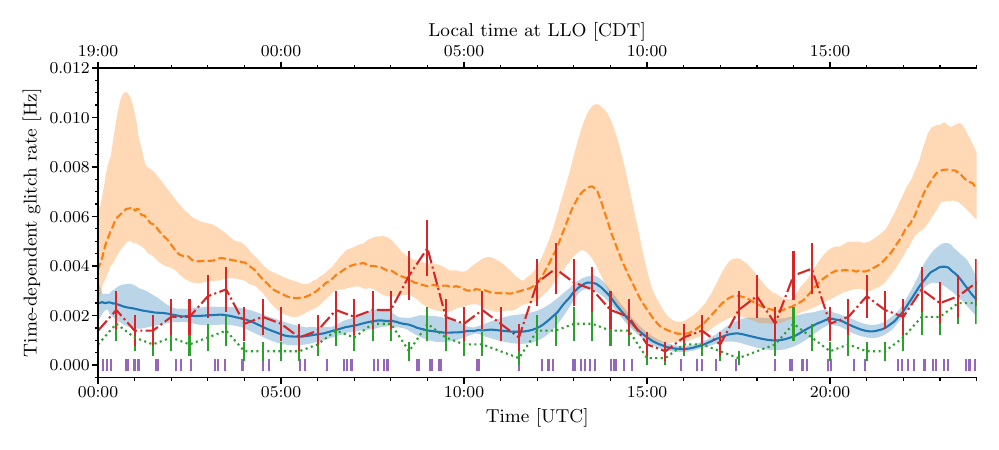}
    \caption{The time-dependent glitch rate inferred from the glitch-only analysis (blue) and glitch and population model (orange) for \qty{1}{\day} of data from \cref{fig:LLO_data}.
    We also show the measured rate inferred using a Poisson binned rate with bin durations of \qty{1}{\hour}, 50\% overlaps, and \OMICRON \ac{SNR} thresholds of \num{6.5} (red) and \num{10} (green).
    Finally, vertical purple markers denote the times of \OMICRON triggers.}
    \label{fig:LLO_dataA_rate_lambdat}
\end{figure*}

In \cref{fig:LLO_dataA_rate_lambdat}, we plot the inferred time-dependence of the glitch rate and compare it to the binned estimate.
We observe two distinct times of elevated glitch activity corresponding to the start and end of the working day at the detector site, when there is a natural increase in traffic and other forms of anthropogenic activity.
While we cannot here causally connect the increased glitch rate to the anthropogenic activity, the coincidence of these features with the working day is suggestive.
\changed{We also caution that the Level-I analysis depends on the \ac{PSD} through the noise-weighted inner product, so a part of the inferred time dependence may reflect the evolution of the detector sensitivity rather than of the physical glitch rate alone; we discuss this systematic in \cref{sec:discussion}.}
Moreover, it is already known that certain glitch types are correlated with anthropogenic activity \citep{Glanzer:2022avx}.
We note that in this work, we cannot distinguish between the rate of different glitch types; only the rate of all glitches is well modelled by \ANTIGLITCH.

Comparing the Level-II analysis to the binned estimate and the times of \OMICRON triggers, such features are not so readily visible as the estimates are noisier.
In part, this is because we model the glitch rate as a smoothly varying function, but it also indicates that the hierarchical model may be able to extract more information from the data than the binned estimate.

\subsection{Compute time}
\label{sec:timing}
The hierarchical model applied to the \ac{LLO} data in this section has distinct advantages in its threshold-free nature and capacity to probe the glitch population as part of the \minorchanged{Level-II} analysis.
However, it is far more computationally demanding than the alternative and standard binned \OMICRON approach.
We discussed this briefly in \cref{sec:levelI}, noting that the dominant factor is the sampling time; there is an additional computational overhead for each analysis of a few percent, which arises from the initialisation (i.e., obtaining the data and \ac{PSD}, and performing post-processing).
Having performed the Level-I analyses for the day of \ac{LLO} data, we can now quantify the compute time and hence the cost of this method in practice.

\begin{figure}
    \centering
    \includegraphics[width=\linewidth]{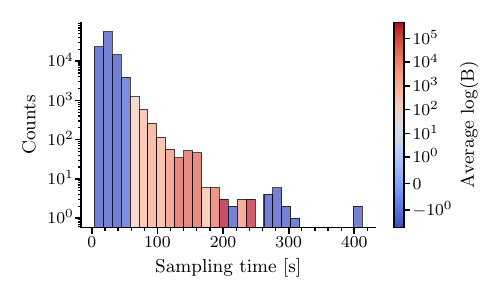}
    \caption{A histogram of the compute time for all Level-I analyses performed for the day of \ac{LLO} data.
    We colour each histogram bin by the average \lnB of analyses within the bin.
    }
    \label{fig:LLO_time}
\end{figure}

In \cref{fig:LLO_time}, we plot a histogram of the sampling time for each Level-I analysis computed as part of this study.
\cref{fig:LLO_time} demonstrates that the majority of analyses take a few tens of seconds, but there is a long tail.
Colouring the histogram bins by the average \lnB within the bin, we can see that segments which contain a glitch (i.e., those with $\lnB > 0$) take longer.
This is expected as we are using nested sampling to compute the evidence and posteriors for which the sampling time scales with the \ac{SNR} \citep{2025PhRvD.112h4039H}.
We additionally find a small secondary peak at a sampling time of around \qty{300}{\second}: we expect this arises from a small number of analyses that landed on a slower node or had a particularly long sampling time for other reasons.

\section{Application: coincident glitch probabilities}
\label{sec:app}

Gravitational-wave search pipelines are the de facto standard for finding and quantifying the significance of candidate astrophysical signals.
While implementation details vary, the general approach is to construct a detection statistic, empirically measure the background distribution, and use this to construct a \ac{FAR}.
The detection statistic typically includes components that weigh in favour of coincident signals between detectors with waveform morphologies consistent with \ac{CBC} signals (for templated searches).
The background distribution is typically calculated using chunks of data and a methodology to produce signal-free background data \citep[see, e.g.][for a description of the methods employed in several pipelines]{LIGOScientific:2025yae}.
This makes the \ac{FAR} highly robust and optimal for detection as it averages over all possible ways that a false alarm could be produced (i.e., glitches, Gaussian noise fluctuations, or other sources).

However, in the case of an ambiguous candidate (e.g., where the \ac{FAR} or other significance estimate lies near typical thresholds used to separate signals from noise), additional investigations may be useful to determine if a candidate is astrophysical in nature. 
To this end, we will now develop a new approach, using the methodology introduced in this paper, to investigate the probability that a candidate is produced by a pair of coincident glitches.

Consider two detectors, which we label \detectorA and \detectorB, each producing glitch triggers as independent Poisson processes with rates \lambdaA and \lambdaB, respectively.
We wish to estimate the probability that the two detectors produce triggers within a time \Tsep\ of one another during an observation of duration \Tspan (over which we assume the rates are constant).
Given a trigger from detector \detectorA, the probability of at least one 
trigger from detector \detectorB falling within a window of $\pm$\Tsep\ is 
$1 - e^{-2\lambdaB\Tsep} \approx 2\lambdaB\Tsep$,
where the approximation holds 
for $2\lambdaB\Tsep \ll 1$, i.e.\ that it is unlikely for more than one glitch from 
detector \detectorB to fall within a single coincidence window. The expected number of 
accidental coincidences over the full observation are then 
$\mu = 2\lambdaA\lambdaB\Tsep\Tspan$, and treating coincidences as Poisson 
distributed, we can calculate the probability of at least one pair of coincident glitches as:
\begin{equation}
    \pcoinc = 1 - e^{-2\lambdaA\lambdaB\Tsep\Tspan}\,.
    \label{eqn:Pexact}
\end{equation}

If we further assume that the expected number of coincidences is small, $2\lambdaA\lambdaB\Tsep\Tspan \ll 1$, then a second expansion gives the approximation
\begin{equation}
    \pcoinc \approx 2\lambdaA\lambdaB\Tsep\Tspan.
    \label{eqn:Papprox}
\end{equation}
Since each rate is estimated from a Poisson count of $n_i = \lambda_i \Tspan$ glitches, standard error propagation on the product in Eq.~\eqref{eqn:Papprox} yields a standard error of
\begin{equation}
    \sigmacoinc = 2\Tsep\sqrt{\Tspan\lambdaA\lambdaB(\lambdaA+\lambdaB)}\,.
    \label{eqn:Perr}
\end{equation}

\cref{eqn:Papprox} and \cref{eqn:Perr} provide a useful approximation to \cref{eqn:Pexact} when the coincident probability is small.
They also reveal the (perhaps expected) result that the probability scales linearly with both the rates and with the separation time \Tsep and data span \Tspan under consideration.
This methodology is not new. For example, a related quantity is used in \citet{Nitz:2017svb} to downweight triggers that occur in noisy data segments.
However, what is novel here is that, for the first time, we have a reliable threshold-free estimate of the rate using the methodology introduced in \cref{sec:methodology}.
The caveat is that this rate is only sensitive to glitches well-modelled by \ANTIGLITCH.

To give an initial quantified example, let us consider the \ac{LHO} and \ac{LLO} detectors, for which the maximum inter-detector travel time for a passing signal is $\sim$\qty{10}{\milli\second}.
Next, let both detectors have a rate similar to the day of data studied in \cref{sec:results}, i.e., we choose $\lambdaA=\lambdaB\approx$\qty{1e-3}{\hertz}.
This is reasonably typical of the best performance the instruments achieve, see, e.g.\ \citep{LIGO:2024kkz}.
Then, over a day, the probability of a pair of coincident glitches is small:
\begin{align}
    \pcoinc \approx 0.002\;
    \left(\frac{\lambdaA}{\qty{0.001}{\hertz}}\right)
    \left(\frac{\lambdaB}{\qty{0.001}{\hertz}}\right)
    \left(\frac{\Tsep}{\qty{10}{\milli\second}}\right)
    \left(\frac{\Tspan}{\qty{1}{\day}}\right)\,.
\end{align}
This demonstrates that coincident glitches are rare for such a glitch rate.
This can be further quantified by considering the expectation value of the waiting time between coincident glitches:
\begin{align}
    \langle t_{\rm CG} \rangle \approx \frac{1}{2\lambdaA\lambdaB\Tsep}
    = \qty{580}{\day}\;
    \left(\frac{\lambdaA}{\qty{0.001}{\hertz}}\right)
    \left(\frac{\lambdaB}{\qty{0.001}{\hertz}}\right)
    \left(\frac{\Tsep}{\qty{10}{\milli\second}}\right)\,.
\end{align}
However, this highlights that the wait time between coincident glitches rapidly decreases if the rates in both detectors increase.

To connect this coincident glitch rate with the methodology developed in \cref{sec:methodology}, we take as an example GW230630\_070659, a \ac{BBH} candidate found by \GSTLAL in O4a \citep{LIGOScientific:2025slb}.
The candidate has a \ac{FAR} of \qty{0.47}{\per\year} and a probability of astrophysical origin of \num{0.88}.
Usually, such statistics would be sufficient to justify considering this as a real astrophysical candidate.
However, evidence was found that the candidate was of instrumental origin.
Specifically, the event validation procedure \changed{\citep{LIGOScientific:2025yae, LIGO:2024kkz}} identified excess power from scattered light in both detectors at the time of the event.

We take \qty{1}{\hour} of data from both \ac{LHO} and \ac{LLO} centred on the trigger time and apply the Level-I analysis to every second of data.
After looking at the initial results, we modified the approach described in \cref{sec:levelI}, increasing the maximum of the $\gamma$ distribution to 1000 as multiple segments had posteriors that hit this upper bound.
Moreover, we also applied a high-pass filter at \qty{15}{\hertz} to the data, as it was found that low-frequency noise was impacting the \ANTIGLITCH analyses.
After applying these changes, we study the individual Level-I analyses and confirm a peak at the trigger time in \ac{LLO}, indicating that \ANTIGLITCH is sensitive to the putative signal. 
For \ac{LHO}, we do not see a peak at the trigger time, but we do see a peak a few seconds after: this is consistent with the data (see \cite{LIGOScientific:2025slb}) where \ac{LHO} is weaker than \ac{LLO} and has a clear artefact a few seconds after.

\begin{figure}
    \centering
    \includegraphics[width=\linewidth]{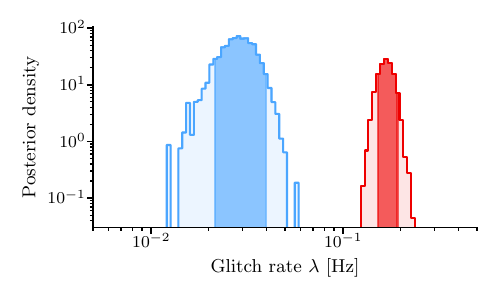}
    \caption{The glitch rate posterior for \ac{LHO} (blue) and \ac{LLO} (red) inferred from \qty{1}{\hour} of data surrounding the candidate GW230630\_070659.
    Shaded regions indicate the 90\% credible interval.
    }
    \label{fig:GW230630_H1L1_rate}
\end{figure}

Taking the Level-I analyses, we then infer the glitch rate for both detectors, jointly with the amplitude distribution.
When modelling the amplitude distribution, we find that if using uninformative priors, the posteriors find a solution with an extreme scaling exponent $\sim -8$ and a glitch rate of \qty{1}{\hertz}, suggesting every segment contains a glitch.
In practice, looking at various data segments, we do observe frequent glitches in both detectors - this is likely connected to the fact that \ANTIGLITCH posteriors favoured large values of $\gamma$.
We also identify that often, the glitches are long in duration, and the \ANTIGLITCH model is a poor fit.
This, coupled with the extreme scaling exponent, indicates the population model is failing.
Therefore, we opt to instead apply priors based on our analysis of the \ac{LLO} data, namely, we fix the amplitude prior to be a power law with a scaling exponent of \num{-1.5} with a minimum of \num{3}.
In \cref{fig:GW230630_H1L1_rate}, we plot the resulting inferred glitch rate for \ac{LHO} and \ac{LLO}.
This figure demonstrates that, even under this restrictive amplitude population prior, the glitch rate is highly elevated relative to the typical behaviour \citep{LIGO:2024kkz}.

\begin{figure}
    \centering
    \includegraphics[width=\linewidth]{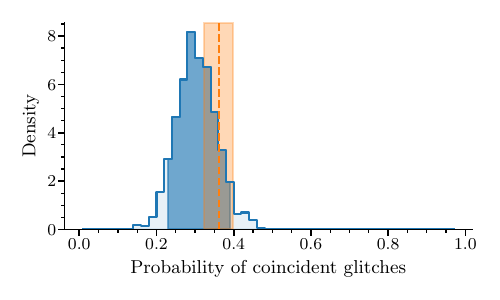}
    \caption{The probability of a pair of coincident glitches \pcoinc, estimated using \cref{eqn:Pexact}, samples from posterior distribution on the rates from \cref{fig:GW230630_H1L1_rate}, $\Tsep=$\qty{10}{\milli\second}, and $\Tspan=$\qty{1}{\hour}.
    We also add the approximation, \cref{eqn:Papprox}, along with its uncertainty.
    }
    \label{fig:GW230630_pcoinc}
\end{figure}

Taking the posterior samples of the glitch rate in \cref{fig:GW230630_H1L1_rate}, in \cref{fig:GW230630_pcoinc}, we then show the distribution of \pcoinc for \qty{1}{\hour}.
This illustrates that the probability of a pair of coincident glitches is large and therefore supports the conclusion in \citet{LIGOScientific:2025slb} that this candidate is likely of terrestrial origin.
\changed{However, this result depends on several analysis choices: the extended prior on $\gamma$, the \qty{15}{\hertz} high-pass filter, and, in particular, the amplitude population fixed by hand after the analysis with uninformative priors failed to produce a sensible solution.
We therefore present this example as a demonstration of how the machinery can be applied to an ambiguous candidate, rather than as an independent determination of the origin of GW230630\_070659.}
For completeness, we also plot the approximation to \pcoinc given by \cref{eqn:Papprox}: in this regime, the approximation is not valid, and it is therefore not too surprising that it does not accurately capture the peak of the full posterior distribution.

\section{Summary and Discussion}
\label{sec:discussion}

In this work, we have introduced a new method for measuring the rate of glitches in interferometric gravitational-wave detectors such as \ac{LIGO}, Virgo, and KAGRA.
Our goal is to provide an estimate of the glitch rate that does not require the application of an arbitrary \ac{SNR} threshold, so that we can smoothly interpolate down to low \ac{SNR}.
\changed{We emphasise that the threshold is not so much eliminated as replaced: in place of a threshold, the method requires a parameterised glitch model, priors on its parameters, and a population model.
Unbiased recovery of the rate is therefore contingent on those choices being adequate, as our sensitivity study in \cref{sec:validation_sensitivity} makes explicit.
Relatedly, the rate we infer is defined with respect to the glitch model adopted in the Level-I analysis.
Throughout this work that model is \ANTIGLITCH, and so the quantity we measure here is the rate of short-duration glitches that \ANTIGLITCH describes well, rather than the total glitch rate of the instrument.
We stress that this is a property of the demonstration rather than of the framework: the mixture likelihood at the heart of the Level-II inference accommodates any glitch model, or several simultaneously, and we return to this below.}

The method utilises hierarchical Bayesian inference.
The Level-I analysis examines \changed{\qty{4}{\second} segments of a given data span, stepped every \qty{1}{\second} and with a \qty{1.1}{\second} central-time prior window,} utilising the \ANTIGLITCH model \citep{Bondarescu:2023jcx}, a simple frequency-domain parameterised glitch model consisting of a Gaussian on the logarithmic frequency.
In the Level-II analysis, we then use the recycling technique to combine the Level-I inferences, providing a means to measure the glitch rate.
We show how this can be done for a simple rate-only estimate and how to model the population properties.
To reduce the computational overhead for the rate and population measurement, we introduce \acf{HIQC}, which uses \Quantiles quantiles from \Msamples posterior samples, producing a speedup of $\Msamples/\Quantiles$.
We validate that with just a handful of quantiles, the approximation error is negligible.
Furthermore, we extend the Level-II analysis to model a time-dependent glitch rate using basis functions and demonstrate that this rate can be jointly inferred alongside the population using \ac{HIQC}.

We perform a validation study on simulated detector data containing Gaussian noise and glitches modelled by \GLITCHFLOW (a normalising flow estimator trained on glitches from the third observing run).
We run a series of studies to validate that the methodology can accurately recover the simulated glitch rate.
We also vary the population properties of the simulated glitches to demonstrate that, provided the hierarchical method correctly models the population properties, it remains unbiased when the distribution of glitches and fluctuations from Gaussian noise heavily overlap.

To understand the performance of the method in practice, we analyse \qty{1}{\day} of continuous data from \ac{LLO} during the fourth observing run.
We find that \ANTIGLITCH finds all glitches identified by \OMICRON except for a small subset of low-frequency long-duration glitches, as expected given that \ANTIGLITCH is known to be a poor model for such morphologies.
From the Level-I analyses, we compute the glitch rate and compare it with estimates based on \OMICRON triggers and a choice of \ac{SNR} threshold.
We find the two methods to produce consistent estimates, with the rate and population inferred glitch rate slightly lower than the rate inferred from all \OMICRON triggers with an \ac{SNR} greater than \num{6.5}.
Next, we infer the time-dependent rate and find that the rate varies throughout the day, with increases of up to a factor of 3 at the beginning and end of the working day.
Such an analysis is more difficult using the binned \OMICRON triggers as the uncertainty on the estimate is large.
Studying the computational cost of the hierarchical method, we find that the cost is dominated by the Level-I analyses, which take around \qty{40}{\second} per \qty{1}{\second} of strain data.
This indicates that, if used at scale (e.g., as a means to measure the glitch rate over an entire observing run), the computational cost would be significant.

Finally, we show how the measured glitch rates can be used to infer the probability of a coincident pair of glitches.
As a worked example, we measure the glitch rate in \qty{1}{\hour} of \ac{LLO} and \ac{LHO} data surrounding GW230630\_070659 and find that the probability is $\approx 0.3$, supporting the conclusion that this event is of terrestrial origin \citep{LIGOScientific:2025slb}.

The hierarchical model developed in this work is novel and provides a unique approach to measuring the glitch rate.
Unlike the standard approach of counting \OMICRON triggers above a threshold, our method does not require an arbitrary threshold and can smoothly interpolate down to low \ac{SNR}.
\changed{
However, in its current form, we do not envision this as a replacement for the \OMICRON approach.
At a cost of $\approx 40$ times real time, it is not a candidate for routine, continuous monitoring of an observing run.
Rather, it is suited to targeted offline studies: validation and follow-up of ambiguous candidates, as demonstrated in \cref{sec:app}; the detailed characterisation of selected intervals, such as periods of unusual detector behaviour or the neighbourhood of a candidate event; and the calibration of threshold-based rates, by establishing how a given \ac{SNR} threshold maps onto the underlying rate.
}
With that in mind, there are a number of avenues for future work.

First, the computational cost of the method is significantly greater than that of the highly optimised \OMICRON pipeline.
This is because the method requires performing a full Bayesian parameter estimation for each segment of data, which is computationally expensive.
We have already reduced the cost by using time and phase marginalisation, but further speed-ups may be possible by using more efficient samplers, gradient-based methods, or by using machine learning techniques such as simulation-based inference \citep{2025arXiv250812939D} to approximate the posterior distributions.
\changed{Such a speed up would aid in targeted follow-up studies of ambiguous candidates, as well as enabling the method to be applied to longer stretches of data.}

Second, in the \minorchanged{Level-II} analysis, we have not fully explored the population inference aspect of the method.
In future work, it would be interesting to explore more complex population models (e.g., involving the frequency, amplitude, and bandwidth of the \ANTIGLITCH model) and include correlations between glitch parameters or allow for multiple sub-populations.

\changed{Third, a systematic that we have not quantified is the evolution of the detector noise, which in real data is known to occur on timescales of hours \citep{LIGO:2024kkz}.
We mitigate this by estimating the \ac{PSD} locally, from the \qty{64}{\second} preceding each segment.
The method is also robust to the overall noise level by construction: the \ac{PSD} enters the Level-I analysis only through the noise-weighted inner product, \cref{eqn:inner_product}, i.e.\ only through the whitened strain, and it does so identically under \modelG and \modelN.
Any \ac{PSD}-dependent normalisation is therefore common to the two evidences and cancels in their ratio, so a change in the overall noise level that the local estimate tracks correctly leaves \lnB unchanged.
However, two effects remain that could bias the inferred rate:
(i) \ac{PSD} estimation is itself noisy, and so the Level-I likelihood is not exactly correct and may be systematically biased; and (ii) the \ANTIGLITCH amplitude $A$ is defined with respect to the noise-weighted inner product and is therefore an \ac{SNR}-like quantity, so a fixed population of physical glitches maps to a distribution in $A$ that shifts as the \ac{PSD} evolves.
Part of the time dependence we infer in \cref{fig:LLO_dataA_rate_lambdat} could therefore reflect changing detector sensitivity rather than a changing physical glitch rate.
This is compounded by our choice to model a time-dependent rate alongside a time-independent population.
A controlled test, in which a known glitch population is injected into off-source data spanning a period of significant \ac{PSD} evolution, would quantify the size of this effect; we regard this as an important limitation of the present validation and leave the test to future work.}

Finally, the current implementation relies on the \ANTIGLITCH model to identify and characterise glitches in the Level-I analysis.
However, the mixture model at the heart of the Level-II inference could be extended to include other glitch models as well to capture a wider variety of glitch morphologies.
Coupling this with clustering techniques to group similar glitches will provide a more comprehensive picture of the glitch population and its properties.
Along similar lines, it may also be interesting to explore a hybrid approach, mapping the information from \OMICRON triggers to Level-I inferences, enabling us to piggyback on the computational efficiency of \OMICRON while still benefiting from the hierarchical inference framework.

In conclusion, we have introduced a novel hierarchical Bayesian method for measuring the glitch rate in interferometric gravitational-wave detectors.
We hope this can spur new thinking in this field and provide a complementary approach to the standard method of counting \OMICRON triggers, enabling us to better understand the glitch population and its impact on gravitational-wave searches.

\section*{Acknowledgements}
\changed{We are thankful to Derek Davis for their comments on the manuscript, which improved the presentation of the results. We are also thankful to the two anonymous referees for their comments, which improved the clarity of the manuscript and corrected several errors in the original version.}

This work is supported by the Science and Technology Facilities Council (STFC) grant UKRI2488.
The authors are grateful for computational resources provided by the LIGO Laboratory and supported by National Science Foundation Grants PHY-0757058 and PHY-0823459.
The authors are also grateful for computational resources provided by Cardiff University and supported by STFC grants ST/I006285/1 and ST/V005618/1.
This research has made use of data or software obtained from the Gravitational Wave Open Science Center (gwosc.org), a service of the LIGO Scientific Collaboration, the Virgo Collaboration, and KAGRA. 
This material is based upon work supported by NSF's LIGO Laboratory which is a major facility fully funded by the National Science Foundation, as well as the Science and Technology Facilities Council (STFC) of the United Kingdom, the Max-Planck-Society (MPS), and the State of Niedersachsen/Germany for support of the construction of Advanced LIGO and construction and operation of the GEO600 detector. 
Additional support for Advanced LIGO was provided by the Australian Research Council. 
Virgo is funded, through the European Gravitational Observatory (EGO), by the French Centre National de Recherche Scientifique (CNRS), the Italian Istituto Nazionale di Fisica Nucleare (INFN), and the Dutch Nikhef, with contributions by institutions from Belgium, Germany, Greece, Hungary, Ireland, Japan, Monaco, Poland, Portugal, and Spain. KAGRA is supported by the Ministry of Education, Culture, Sports, Science and Technology (MEXT), Japan Society for the Promotion of Science (JSPS) in Japan; the National Research Foundation (NRF) and Ministry of Science and ICT (MSIT) in Korea; Academia Sinica (AS) and National Science and Technology Council (NSTC) in Taiwan.

We utilise the \NUMPY \citep{harris2020numpy} and \SCIPY library \citep{2020SciPy-NMeth} for data processing and analysis, and we also use the \MATPLOTLIB \citep{Hunter:2007ouj} library for visualisation.

\section*{Data Availability}
The interferometric strain data used in this work is publicly available from the Gravitational Wave Open Science Center (\href{https://www.gwosc.org/}{www.gwosc.org}).
For reproducibility of results within this work, see the data release \citep{ashton_2026_19630780} which contains software used to create simulated data and perform all analyses.

\bibliographystyle{mnras}
\bibliography{references}

\appendix
\section{Missed triggers}
\label{app:examples}
\begin{figure*}
    \centering
    \includegraphics[width=\linewidth]{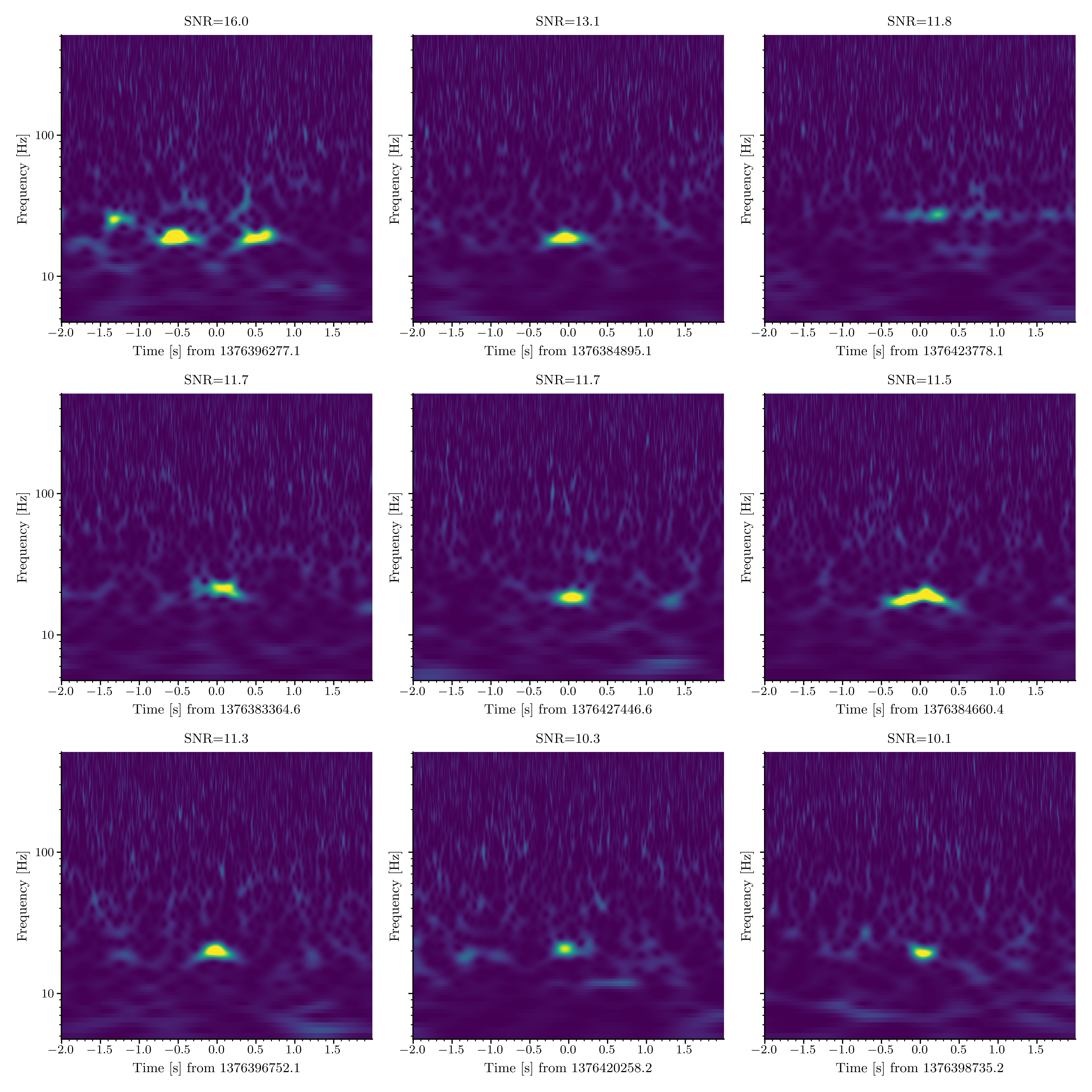}
    \caption{Time-frequency spectrograms of \ac{LLO} data segments in our study when the \OMICRON trigger \ac{SNR} is greater than 10, but \lnB from the \minorchanged{Level-I} analysis is negative, indicating the \minorchanged{\ANTIGLITCH} model did not find evidence for a signal.}
    \label{fig:LLO_dataA_omega_scans}
\end{figure*}

In \cref{fig:LLO_dataA_omega_scans}, we provide time-frequency spectrograms of the \ac{LLO} data missed by our analysis in \cref{sec:results}.

\bsp
\label{lastpage}
\end{document}